\documentclass{aa}
\usepackage{txfonts}
\usepackage{epsfig}
\usepackage{subfigure}

\usepackage{times}
\usepackage{natbib}
\bibpunct{(}{)}{;}{a}{}{,}

\begin{document}

\title{X-ray reprocessing in Seyfert Galaxies: simultaneous XMM-\textit{Newton}/BeppoSAX observations}
\author{S. Bianchi\inst{1}, G. Matt\inst{1}, I. Balestra\inst{1}, M.
Guainazzi\inst{2}, G.C. Perola\inst{1}}

\offprints{Stefano Bianchi\\ \email{bianchi@fis.uniroma3.it}}

\institute{Dipartimento di Fisica, Universit\`a degli Studi Roma Tre, Italy
\and XMM-Newton Science Operation Center/RSSD-ESA, Villafranca del Castillo,
Spain }

\date{Received / Accepted}

\authorrunning{S. Bianchi et al.}

\abstract{We selected a sample of eight bright unobscured (at least at the iron line energy) Seyfert Galaxies observed simultaneously by XMM-\textit{Newton} and BeppoSAX, taking advantage of the complementary characteristics of the two missions. The main results of our analysis can be summarized as follows: narrow neutral iron lines are confirmed to be an ubiquitous component in Seyfert spectra; none of the analyzed sources shows unambiguously a broad relativistic iron line; all the sources of our sample (with a single exception) show the presence of a Compton reflection component; emission lines from ionized iron are observed in some sources; peculiar weak features around 5-6 keV (possibly arising from rotating spots on the accretion disk) are detected in two sources. The scenario emerging from these results strongly requires some corrections for the classical model of reprocessing from the accretion disk. As for materials farther away from the Black Hole, our results represent a positive test for the Unification Model, suggesting the presence of the torus in (almost) all sources, even if unobscured.


}

\maketitle

\section{Introduction}

The X-ray spectrum of Seyfert Galaxies consists of several components, whose relative importance changes with energy. It is therefore important to observe them in an energy range as wide as possible. In this respect, simultaneous XMM-\textit{Newton}/BeppoSAX observations represent a unique opportunity to analyze the relations between different parts of the spectrum. Indeed, the combination of XMM-\textit{Newton} superior effective area and BeppoSAX broad band coverage makes this kind of observations a powerful tool to disentangle the various components and then put them in the context of the overall spectrum of Seyfert Galaxies.

In particular, simultaneous observations are ideally suited for the study of the origin of the iron lines. On one hand, with XMM-\textit{Newton} one can study in detail the line profile, which provides valuable hints on the distance of the emitting matter from the Black Hole (BH) by dynamic arguments. If the line is produced in the innermost regions of the accretion disk, the resulting profile has a characteristic double-peaked shape, due to Doppler and gravitational effects \citep[see][for a review]{fab00}. If, instead, the line is produced much farther away from the BH, as for instance in the `torus' invoked in Unification Models \citep{antonucci93}, a symmetric profile is preserved, broadening by kinetic motion of the emitting gas being likely below spectral resolution of present detectors. Finally, if the line is produced in the Broad Line Region (BLR), its width should be consistent with that of the broad optical/UV lines. On the other hand, BeppoSAX has proven so far to be the best X-ray mission to measure the amount of the Compton reflection component, which can be crucial to discern between an origin of the line from Compton-thick or Compton-thin matter \citep{mgm03}.

Simultaneous XMM-\textit{Newton}/BeppoSAX observations of Seyfert Galaxies have already been analyzed in the past: NGC~5506 \citep{matt01,bianchi03}, NGC~7213 \citep{bianchi03b}, NGC~5548 \citep{pounds03}, IC~4329A \citep{gond01}, MKN~841 \citep{petr02} and MCG-6-30-15 \citep{fab02}. In this paper, we selected a sample of eight sources with simultaneous observations, with the aim to understand the origin of their iron lines.

\section{Observations and data reduction}

\subsection{The sample}

We searched the archives for all the Seyfert Galaxies observed simultaneously by XMM-\textit{Newton} and BeppoSAX. Among these sources, we decided to exclude some objects, for different reasons: MCG-6-30-15 has a very complex spectrum and iron line, making it probably an exception among Seyfert 1s \citep[see e.g.][ for the analysis of the simultaneous observation]{bvf03}; in the case of MKN~766, a BL LAC object is present in the BeppoSAX/PDS field of view, whose contribution to the total spectrum is impossible to determine precisely because lies outside the XMM-\textit{Newton} and BeppoSAX/MECS field of view \citep{matt00}; NGC~4151 is highly variable and probably observed through a very complex partial absorber \citep[see e.g.][]{sw02}.

After this process of selection, we ended up with eight sources, whose analyzed observations are listed on Table \ref{log}. The resulting sample includes 6 Seyfert 1s and two Compton-thin sources (NGC~5506 and NGC~7213). Three of these simultaneous observations, namely those relative to ESO198-G024, MCG-02-58-022 and NGC~3516, are yet unpublished: when this is the case, we defer the reader to the individual notes on these sources for the references where only the XMM-\textit{Newton} or BeppoSAX observation was analyzed separately. As for the remaining sources, we re-analyzed all of them starting from the reduction process, as described below. This choice was necessary to guarantee homogeneity in data reduction and analysis.

\begin{table*}

\caption{\label{log}The log of all the analyzed observations. The reported fluxes are relative to the fits shown in Table \ref{fitnosoft}.}
\begin{center}
\begin{tabular}{cccccccccc}
\textbf{Object} & \textbf{Date} &\multicolumn{2}{c}{\textbf{EPIC-pn}}& \textbf{PDS} & \textbf{PDS/pn} & \textbf{z} & $\mathbf{N_\mathrm{Hg}}^{c}$ & \textbf{F}$_\mathrm{2-10\,keV}$ & $Log{L_\mathrm{2-10\,keV}}^{d}$\\
&& \textbf{Mode}$^{a}$ & \textbf{T$^{b}$ (ks)} & \textbf{T (ks)} &\textbf{factor}& & $\mathbf{10^{20}}$ \textbf{cm}$\mathbf{^{-2}}$ & $\times10^{-11}$ (cgs) & \\
\hline \textbf{ESO198-G024} & 2001-01-23/26 & M/F &16 & 63 & 0.99 & 0.045500 &3.09 & 1.3 & 43.77\\
\hline \textbf{MCG-02-58-022} & 2000-12-01/05 & T/S & 7 & 47 & 0.92 & 0.047316 & 3.60 & 3.1 & 44.18\\
\hline \textbf{NGC 7213} & 2001-05-27/30 & T/S & 30 &38 & 0.89 &0.005977 & 2.04 & 2.2 & 42.24\\
\hline \textbf{NGC 5548} & 2001-07-08/12 & T/S & 49 & 48 & 0.91 &0.017175 & 1.69 & 4.0 & 43.41\\
\hline \textbf{MKN 841} & 2001-01-11/14 & T/S & 16 & 41 & 0.90 &0.036422 & 2.34 & 1.6 & 43.67\\
\hline \textbf{IC 4329A} & 2001-01-31/02-01 & M/F & 10 &16 & 1.01 &0.016054 & 4.42 & 3.7 & 43.39\\
\hline \textbf{NGC 3516} & 2001-04-10/12 & T/S & 57 & 40 & 1.08 &0.00884 & 3.05 & 2.4 & 42.69\\
\hline \textbf{NGC 5506} & 2001-02-01/03 & M/L & 14 & 38 & 1.16 &0.006181 & 3.81 & 7.8 & 42.95\\
\hline
\end{tabular}
\end{center}

$^{a}$ Filter/Frame: M=Medium, T=Thin, F=Full Frame, L=Large Window, S=Small Window.\\
$^{b}$ After screening for intervals of flaring particle background (see text for details).\\
$^{c}$ Values from \citet{dl90}.\\
$^{d}$ Log of luminosity corrected for absorption (see Table \ref{fitnosoft}) in erg s$^{-1}$ (H$_\mathrm0$=70 km s$^{-1}$ Mpc$^{-1}$)

\end{table*}

\subsection{\label{xmmdata}XMM-\textit{Newton}}

In all cases, the selected science modes produced EPIC pn \citep{struder01} count rates well below the maximum for 1\% pileup (see Table 3 of the XMM-\textit{Newton} Users' Handbook), with the only exception of IC~4329A, whose reduction will be described in detail below. Pileup problems are instead more common in MOS data, so we conservatively decided not to use this instrument and to limit our analysis to X-ray events corresponding to pattern 0 for the pn, in order to have homogeneous choices for the whole sample. All datasets have been reduced with \textsc{SAS} 5.4.1 and screened for intervals of flaring particle background, by inspecting the high energy (E $>$ 10 keV) light curve \citep{lumb02}. An extraction radius of 40$\arcsec$ was chosen for pn spectra and lightcurves.

Following \citet{gond01}, we decided to mitigate the pileup problems of IC~4329A by excluding photons from the core of the image. We therefore extracted spectra from annuli with outer radius of 80$\arcsec$ and increasing inner radii, performing spectral fits until the parameters (in particular the absorbing column density and the power law index) reached stable values. This process ended in the choice of an inner radius of 20$\arcsec$, very similar to the one used by \citet{gond01}.

Finally, some words must be spent on the data reduction of MKN~841. The XMM-\textit{Newton} observation of this source was split in two parts, due to operational needs. However, the two segments are separated by only 15 hours and were performed with the same combination of array and filter. Furthermore, the flux variation between the two parts is lower than 10\% and is not accompanied by a significant variation of the spectral parameters. Therefore, we decided to obtain a single event file using the \textsc{SAS} tool \textsc{merge} and then extract a combined spectrum: the exposure time listed in Table \ref{log} is the total time after this operation.

\subsection{BeppoSAX}

MECS \citep{boellamecs97} and PDS \citep{fronterapds97} spectra were downloaded from the ASDC archive\footnote{http://asdc.asi.it/}. Since all PDS spectra were extracted with fixed rise time threshold, we adopted a normalization factor of 0.86 between the PDS and the MECS \citep{fiore99}. This factor was then rescaled to the one found between the pn and the MECS spectra in a combined fit. As a result of this process, we eventually obtained the normalization factors between the EPIC pn and the PDS for each source: they are listed in Table \ref{log} \citep[see also][for details on this procedure]{matt01,bianchi03,bianchi03b}.

\subsection{\label{chandradata}\textit{Chandra}}

We also analyzed \textit{Chandra} HETG observations of the sources, when available, to gain further insights on the iron line properties. IC~4329A was observed by \textit{Chandra} ACIS-S HETG on August 26 2001 for 59 ks. NGC~3516 was observed by \textit{Chandra} ACIS-S HETG for 73 ks between April 10 and 11 2003, thus overlapping the simultaneous XMM-\textit{Newton}/BeppoSAX observation analyzed in this paper. Data were reduced with the Chandra Interactive Analysis of Observations software (\textsc{CIAO} 3.0), using the Chandra Calibration Database (\textsc{CALDB} 2.23).\\

All spectra were analyzed with \textsc{Xspec} 11.2.0. In the following, errors correspond to the 90\% confidence level for one interesting parameter ($\Delta \chi^2 =2.71$), where not otherwise stated.

\section{Data analysis}

\subsection{\label{variab}Variability}

Before the spectral analysis, we examined the lightcurves and hardness ratios from EPIC pn and PDS data. No significant flux variability was found for ESO198-G024, MCG-02-58-022, NGC~7213 and IC~4329A, either in the pn or in the PDS observations. As already mentioned, the 10\% flux variability between the two nearly contiguous pn observations of MKN~841 is not accompanied by spectral variations. A flux variability of the order of 20\% around the average value was instead found in the pn observations of NGC~5548 and NGC~5506, but neither source shows also spectral variations. Finally, NGC~3516 reveals a complex temporal behaviour, but no significant spectral variability is apparent.

We also checked for flux variations of the neutral iron line during the EPIC pn observations, extracting lightcurves in the band 6-6.6 keV and studying the hardness ratios against the 3-5 keV and 7-10 keV lightcurves. In all cases, no significant variation is found for the iron line flux. The only exception is possibly MKN~841, as already reported in detail by \citet{petr02}.

\subsection{\label{models}The models}

The adopted baseline model spectrum (BMS) for all sources is very similar to the one used by \citet{per02} for a sample of Seyfert 1s observed by BeppoSAX. In detail, our BMS consists of: a power law with exponential cutoff, together with reflection from an isotropically illuminated cold slab, with an inclination angle relative to the line of sight fixed to 30$\degr$ and subtending the solid angle $R=\Omega/2\pi$ \citep[model \textsc{pexrav} on \textsc{Xspec}:][]{mz95}; cold absorption from the Galactic hydrogen column density, as reported in Table \ref{log}; further absorption from neutral gas at the redshift of the source; a Gaussian line to reproduce the iron K$\alpha$ fluorescent line at the redshift of the source, with energy, width and intensity as free parameters; a further Gaussian line to reproduce the iron K$\beta$, with energy fixed at 7.06 keV (rest-frame), width set equal to that of the K$\alpha$ and intensity left as the only free parameter. Further Gaussian lines, when required by the data, were added to the fits of single sources. When referring to `unresolved' Gaussian lines, we mean $\sigma=1$ eV, which is much lower than the EPIC pn resolution.

When no absorption in excess of the Galactic one is found, the fits were performed with absorption frozen to the Galactic value (listed in Table \ref{log})  and data below 2.5 keV were ignored in all the following analysis. This choice reflects the main objectives of this paper, which focuses on the iron lines and the Compton reflection component, while avoiding, where possible, the complexities of the soft X-ray spectra. On the other hand, this simplification was not applicable for the three sources (IC~4329A, NGC~3516 and NGC~5506) which are absorbed by a significant column density of neutral or ionized gas. In particular, since in these cases a low energy cutoff at 2.5 keV could cause a wrong determination of the column density (and, consequently, of $\Gamma$ and R), we decided to keep the soft X-ray spectrum in the fits, adopting a 0.5-220 keV band for these sources only. Of course, this choice leads to the inclusion of additional components to the BMS, in order to get a good fit to the data below 2.5 keV. We defer the reader to the individual notes on these sources for details. Finally, a 15-220 keV band was always chosen for the PDS.

The best fit parameters of the BMS for all the sources are reported in Table \ref{fitnosoft}. The values of the $\chi^2$/dof are all acceptable, the worst being the ones performed on the broad band, where the complexity in the low energy components greatly affects the quality of the fit. To check for the presence of a further component to the iron line, we added a relativistic line profile \citep[model \textsc{diskline} in \textsc{Xspec}: ][]{fab89} to the BMS. We fixed the inner radius to 6 r$_\mathrm{g}$, the index of the emissivity law $\beta$ to -2 and the line energy to 6.4 keV (rest frame), while keeping all the parameters of the narrow\footnote{We will use the term `narrow' to indicate line widths much smaller than those arising from broadening effects in the accretion disk. In this sense, we will refer to narrow lines even if their widths are resolved.} Gaussian line frozen to those of the previous fit. Furthermore, the inclination angle of the \textsc{diskline} was linked to that of the \textsc{pexrav}, as expected if both the iron line and the Compton reflection component are produced by the accretion disk. The EWs of the relativistic lines resulted always in quite tight upper limits, which are also reported in Table \ref{fitnosoft}. A similar fit was performed adding to the BMS a \textsc{laor} iron line model \citep{laor91}, which takes into account emission from the inner radii of a fully rotating BH, up to the last stable orbit for this case, that is 1.23 r$_\mathrm{g}$. Again, we fixed $\beta=-2$, E=6.4 keV and the narrow line parameters at the values found with the BMS. The $\chi^2$ obtained with this model are statistically equivalent to that of the BMS for most sources and better in some cases (Table \ref{laor}). An alternative approach consisted in the modification of the BMS, replacing the narrow Gaussian line with a \textsc{diskline} relativistic line profile, parameterized as described above. The fit for this model is statistically equivalent to the BMS in all the sources (see Table \ref{fitdiskline}). We will discuss the implications of these results in Sect. \ref{discussion}.\\

\begin{table*}

\caption{\label{fitnosoft}Best fit parameters for the BMS. The second-last column refers to the addition of a \textsc{diskline} component to the model (see text for details). Note that the BMS includes, in some cases, additional features when required by the data: see the analysis of individual sources for details.}
\begin{center}
\begin{tabular}{cccccccccc}
\textbf{Object} &\textbf{N}$_\mathrm{H}$ & $\Gamma$ & \textbf{R} & \textbf{E}$_\mathrm{c}$& E$_\mathrm{Fe}$& \textbf{EW} & $\sigma$ & disk EW &$\chi^{2}$/dof\\
 & $\times10^{22}$ cm$^{-2}$ & & & (keV) & (keV) &(eV) & (eV) &(eV)&\\
\hline \textbf{ESO198-G024}  & - & $1.91^{+0.11}_{-0.07}$ & $0.86^{+0.41}_{-0.64}$ & $>260$ & $6.38\pm0.06$ & $57^{+30}_{-29}$ & $<160$ & $<150$ &103/123\\
\hline \textbf{MCG-02-58-022}  & - &$1.72^{+0.08}_{-0.06}$ & $0.41^{+0.27}_{-0.30}$ &$150^{+300}_{-50}$   & $6.29^{+0.26}_{-0.06}$ & $45^{+85}_{-24}$ & $<340$ & $<140$ & 127/131\\
\hline \textbf{NGC 7213} & - & $1.68^{+0.03}_{-0.02}$ & $<0.19$ & $90^{+50}_{-20}$ & $6.39^{+0.01}_{-0.02}$  &$82^{+16}_{-14}$ & $<60$ & $<20$ &174/177\\
\hline \textbf{NGC 5548} & - & $1.68\pm0.03$ & $0.45^{+0.20}_{-0.16}$ & $190^{+140}_{-60}$ & $6.38^{+0.02}_{-0.03}$ &$58^{+15}_{-14}$ &$63^{+45}_{-58}$ & $<40$ &206/192\\
\hline \textbf{MKN 841} & - & $1.97^{+0.11}_{-0.10}$ & $1.31^{+1.21}_{-0.70}$ & $>100$ & $6.35\pm0.05$ &$96^{+44}_{-35}$ &$88^{+60}_{-53}$ & $<40$ &107/134\\
\hline \textbf{IC 4329A} & $0.26\pm0.01$ & $1.78\pm0.01$ & $0.30^{+0.05}_{-0.06}$ & $130\pm10$ & $6.42\pm0.05$ & $127^{+32}_{-37}$ & $108^{+53}_{-37}$ & $<50$ &263/207\\
\hline \textbf{NGC 3516}  & $4.3^{+0.2}_{-0.1}\times 10^{-2}$ &$1.56\pm0.01$ & $0.50^{+0.08}_{-0.07}$ &$120\pm20$ & $6.40^{+0.01}_{-0.02}$ & $106^{+15}_{-10}$ &$40^{+20}_{-23}$ & $<20$ &296/256\\
\hline \textbf{NGC 5506} & $3.02^{+0.04}_{-0.05}$ & $1.88^{+0.05}_{-0.02}$ &$1.10^{+0.19}_{-0.14}$ &$140^{+40}_{-30}$ & $6.41\pm0.02$ & $100\pm19$ & $59^{+32}_{-33}$ & $<40$ &297/255\\
\hline
\end{tabular}
\end{center}

\end{table*}

\begin{table*}

\caption{\label{laor}Best fit parameters when a \textsc{laor} component is added to the BMS, with the emissivity law $\beta$ fixed to -2. The values for $\beta$ and $\chi^{2}_{\beta}$ refers to the fit where the $r_\mathrm{out}$ parameter is fixed to 400 r$\mathrm{_g}$ (see text for details).}
\begin{center}
\begin{tabular}{cccccccccccc}
\textbf{Object} &\textbf{N}$_\mathrm{H}$ & $\Gamma$ & \textbf{R} & \textbf{E}$_\mathrm{c}$& \textbf{EW} & r$\mathrm{_{out}}$ & $\chi^{2}$/dof & $\beta$ & $\chi^{2}_{\beta}$/dof \\
& $\times10^{22}$ cm$^{-2}$ & & & (keV) &(eV) & (r$\mathrm{_g}$) &&&\\
\hline \textbf{ESO198-G024} & - & $1.95^{+0.12}_{-0.10}$ & $1.10^{+0.98}_{-0.56}$ & $>260$ & $210^{+116}_{-124}$ & $4.6^{+2.3}_{-0.7}$ & 96/124 & $4.5^{+1.8}_{-1.1}$ & 97/124\\
\hline \textbf{MCG-02-58-022} & - & $1.75^{+0.08}_{-0.07}$ & $0.40^{+0.37}_{-0.26}$ & $230^{+520}_{-110}$ & $90^{+92}_{-72}$ & $>22$ & 123/132 & $<3$ & 123/132\\
\hline \textbf{NGC 7213} & - & $1.68^{+0.04}_{-0.06}$ & $<0.55$ & $100^{+290}_{-50}$ & $111^{+39}_{-66}$ & $3.1\pm0.2$ & 161/178 & $6.5^{+1.4}_{-0.9}$ & 163/178\\
\hline \textbf{NGC 5548} & - & $1.66^{+0.03}_{-0.05}$ & $0.48^{+0.21}_{-0.14}$ & $160^{+90}_{-40}$ & $66^{+23}_{-31}$ & $2.7^{+0.2}_{-0.1}$ & 194/193 & $6.8^{+0.9}_{-1.1}$ & 188/193\\
\hline \textbf{MKN 841} & - & $1.86^{+0.14}_{-0.12}$ & $1.49^{+0.75}_{-0.90}$ & $80^{+290}_{-40}$ &$158^{+77}_{-83}$ & $2.3^{+0.2}_{-0.1}$ & 97/135 & $7.8\pm1.2$ & 99/135\\
\hline \textbf{IC 4329A} & $0.27\pm0.01$ & $1.84\pm0.01$ & $0.50\pm0.06$ & $170\pm20$ & $298^{+102}_{-87}$ & $5.5\pm0.1$ & 248/208 & $3.6^{+1.5}_{-0.2}$ & 265/208\\
\hline \textbf{NGC 3516} & $2.2^{+0.4}_{-0.3}\times 10^{-2}$ & $1.49^{+0.02}_{-0.04}$ & $0.53^{+0.17}_{-0.14}$ & $100^{+30}_{-20}$ & $209^{+43}_{-49}$ & $4.8^{+0.8}_{-0.6}$ & 269/257 & $4.8^{+0.9}_{-0.4}$ & 260/257\\
\hline \textbf{NGC 5506} & $2.91^{+0.04}_{-0.05}$ & $1.83^{+0.08}_{-0.04}$ & $1.04^{+0.20}_{-0.16}$ & $110\pm20$ & $124^{+56}_{-49}$ & $2.7^{+0.3}_{-0.1}$ &286/256 & $6.2^{+1.5}_{-1.3}$ & 290/256\\
\hline
\end{tabular}
\end{center}

\end{table*}

\begin{table*}

\caption{\label{fitdiskline}Best fit parameters when the BMS is modified by replacing the narrow iron line with a \textsc{diskline} (see text for details).}
\begin{center}
\begin{tabular}{ccccccccccc}
\textbf{Object} &\textbf{N}$_\mathrm{H}$ & $\Gamma$ & \textbf{R} & \textbf{E}$_\mathrm{c}$& E$_\mathrm{Fe}$& \textbf{EW} & r$\mathrm{_{out}}$ & $\chi^{2}$/dof\\
& $\times10^{22}$ cm$^{-2}$ & & & (keV) & (keV) &(eV) & (r$\mathrm{_g}$) &\\
\hline \textbf{ESO198-G024} & - & $1.91^{+0.04}_{-0.03}$ & $0.86^{+0.83}_{-0.30}$ & $>350$ & $6.4^{+0.8}_{-0.5}$ & $71^{+34}_{-39}$ & $>3000$ & 106/123\\
\hline \textbf{MCG-02-58-022} & - & $1.74^{+0.08}_{-0.06}$ & $0.38^{+0.38}_{-0.26}$ & $220^{+780}_{-100}$ & $6.39^{+0.09}_{-0.13}$ & $131^{+63}_{-60}$ & $>170$ & 122/131 \\
\hline \textbf{NGC 7213} & - & $1.68^{+0.05}_{-0.04}$ & $<0.32$ & $100^{+300}_{-50}$ & $6.39^{+0.02}_{-0.01}$ & $100^{+19}_{-16}$ & $>4\times10^5$ & 178/177\\
\hline \textbf{NGC 5548} & - & $1.68^{+0.03}_{-0.02}$ & $0.45^{+0.19}_{-0.16}$ & $200^{+80}_{-70}$ & $6.37^{+0.03}_{-0.02}$ & $67^{+13}_{-15}$ & $>20\,000$ & 205/192\\
\hline \textbf{MKN 841} & - & $1.96^{+0.04}_{-0.08}$ & $1.25^{+0.31}_{-0.69}$ & $170^{+370}_{-70}$ & $6.35\pm0.06$ & $108^{+40}_{-38}$ & $>3\,500$ & 110/134\\
\hline \textbf{IC 4329A} & $0.26\pm0.01$ & $1.78\pm0.01$ & $0.30^{+0.04}_{-0.06}$ & $130\pm15$ & $6.38^{+0.09}_{-0.03}$ & $193^{+26}_{-84}$ & $>800$ & 266/207\\
\hline \textbf{NGC 3516} & $4.4\pm0.2\times 10^{-2}$ & $1.56\pm0.01$ & $0.47\pm0.07$ & $120\pm20$ & $6.40^{+0.01}_{-0.02}$ & $131\pm16$ & $>1\times10^6$ & 303/256\\
\hline \textbf{NGC 5506} & $3.02^{+0.03}_{-0.06}$ & $1.88^{+0.05}_{-0.04}$ & $1.09^{+0.20}_{-0.13}$ &$140^{+40}_{-30}$ & $6.40^{+0.03}_{-0.01}$ & $117^{+20}_{-22}$ & $>24\,000$ &299/255\\
\hline
\end{tabular}
\end{center}

\end{table*}

\begin{figure*}
\begin{minipage}[b]{0.5\textwidth}
\begin{center}
\epsfig{file=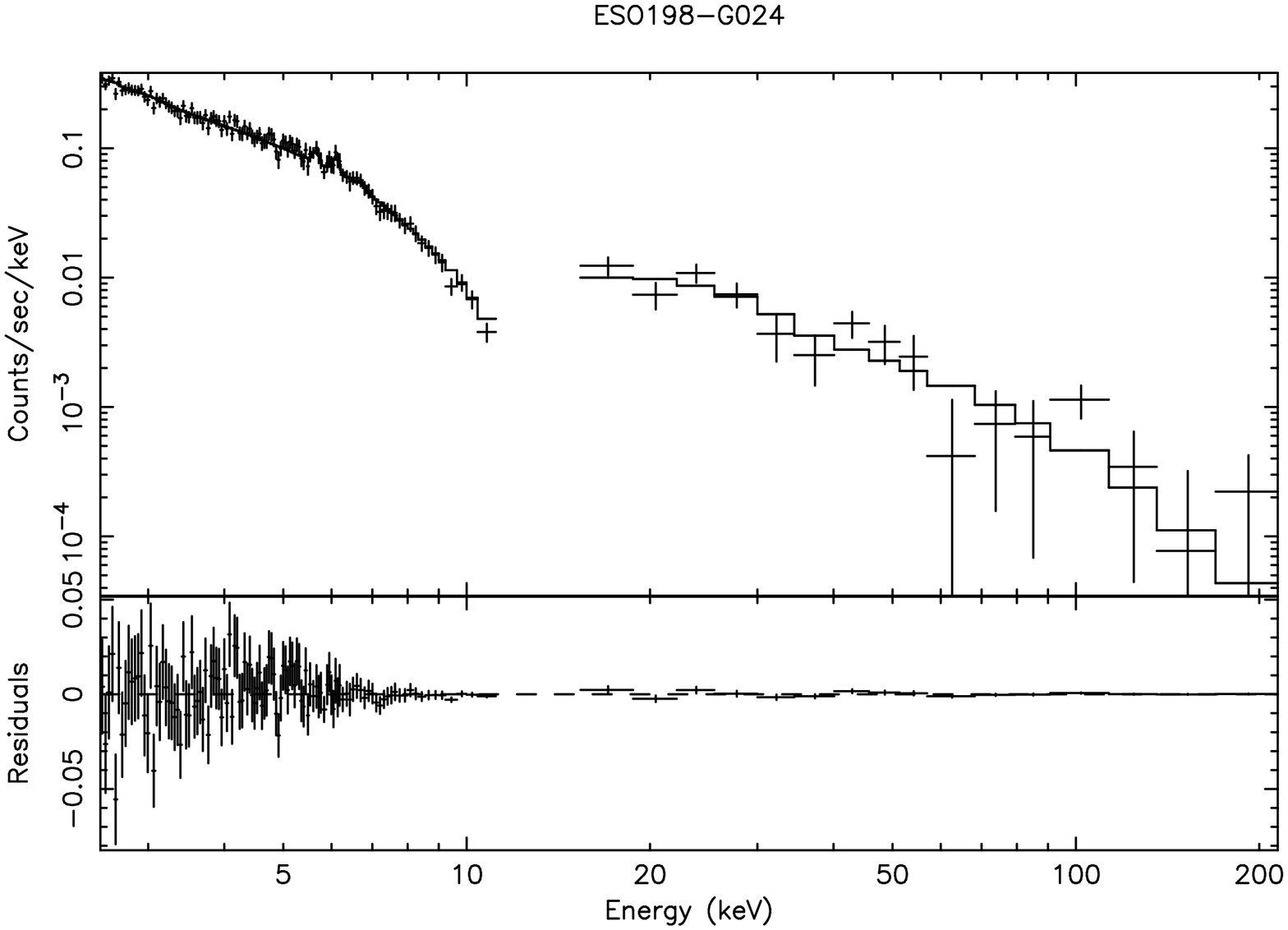,width=5.6cm,angle=-90}
\end{center}
\end{minipage}
\medskip
\begin{minipage}[b]{0.5\textwidth}
\begin{center}
\epsfig{file=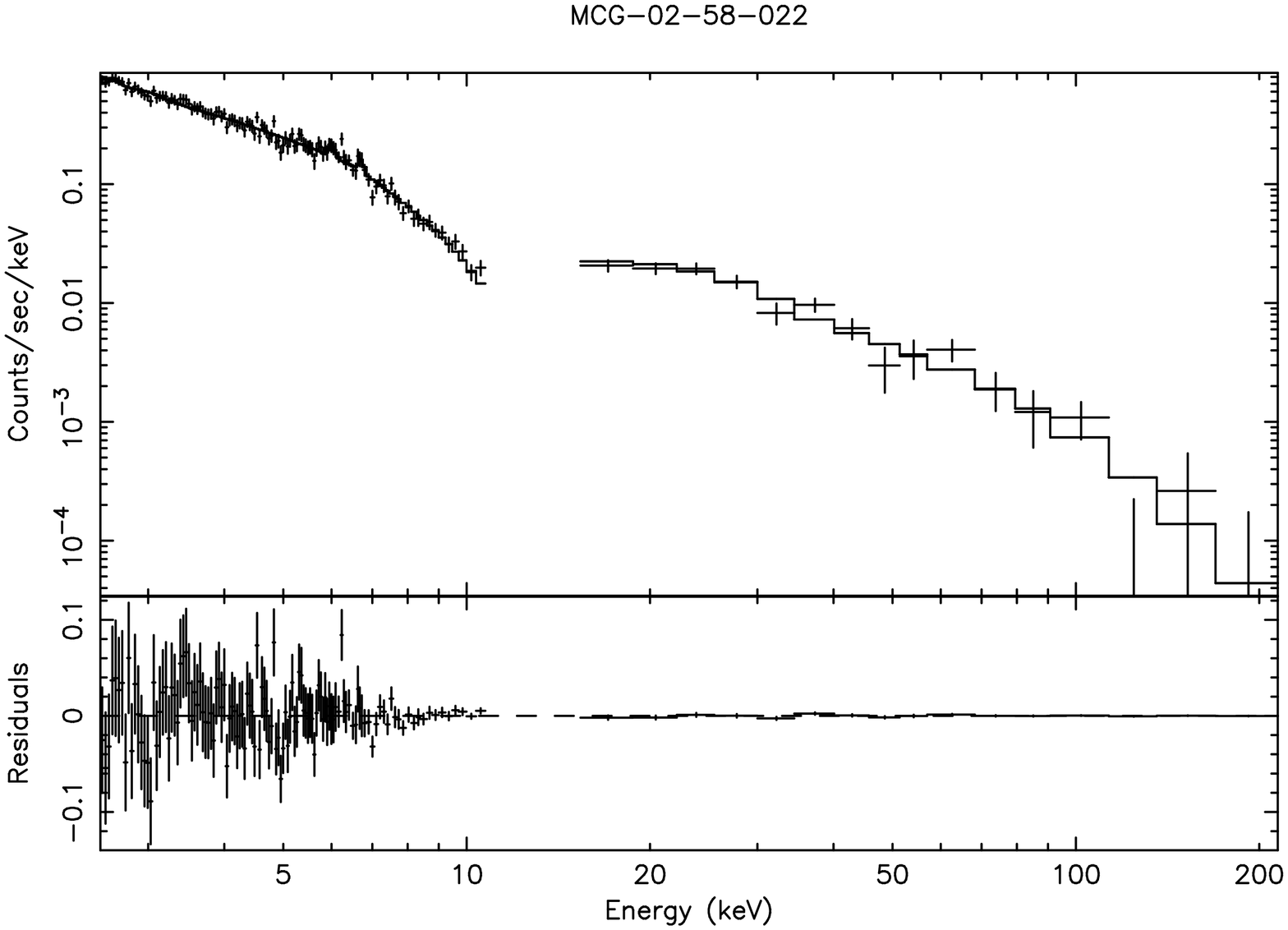,width=5.6cm,angle=-90}
\end{center}
\end{minipage}
\medskip
\begin{minipage}[b]{0.5\textwidth}
\begin{center}
\epsfig{file=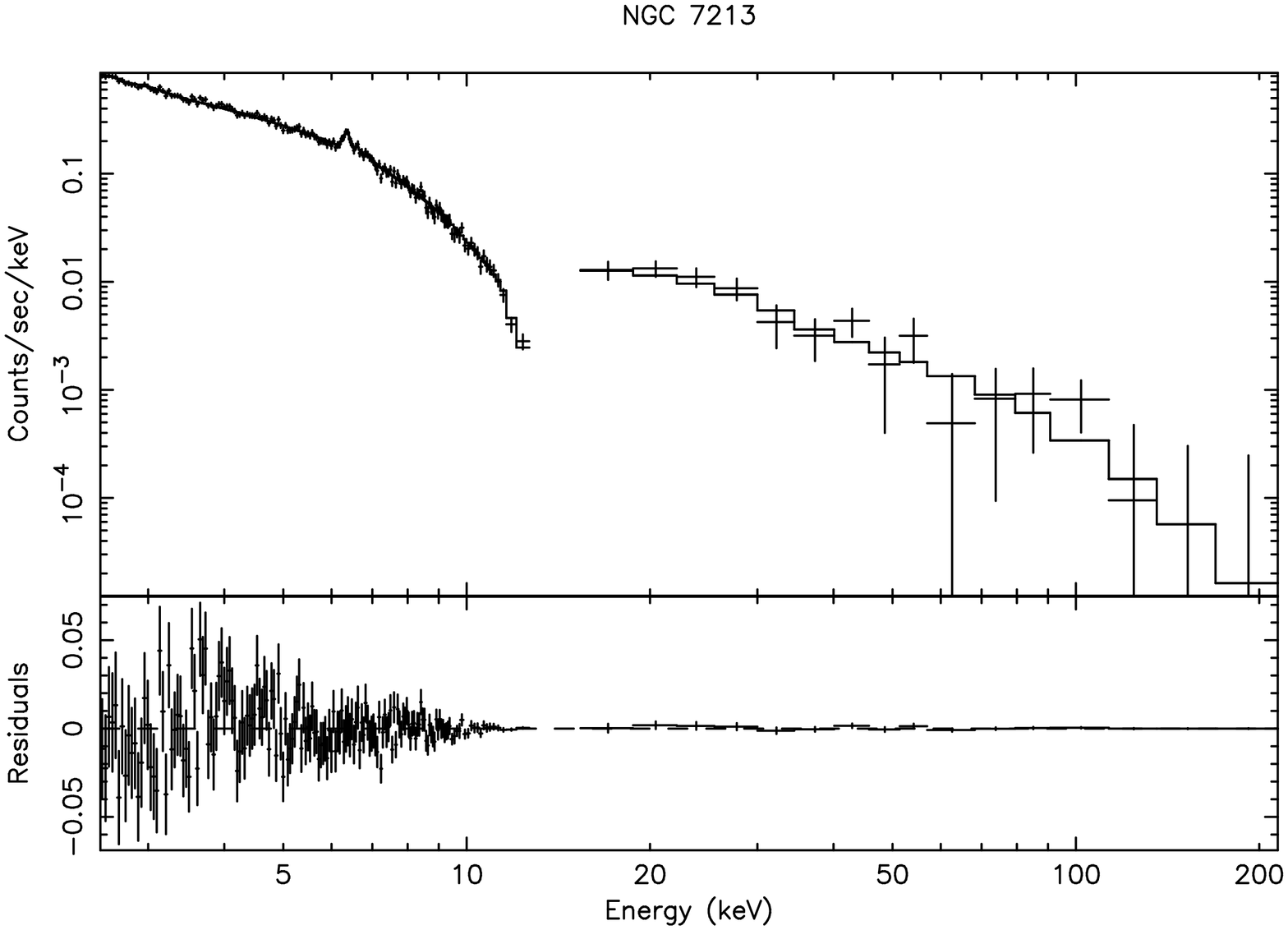,width=5.6cm,angle=-90}
\end{center}
\end{minipage}
\medskip
\begin{minipage}[b]{0.5\textwidth}
\begin{center}
\epsfig{file=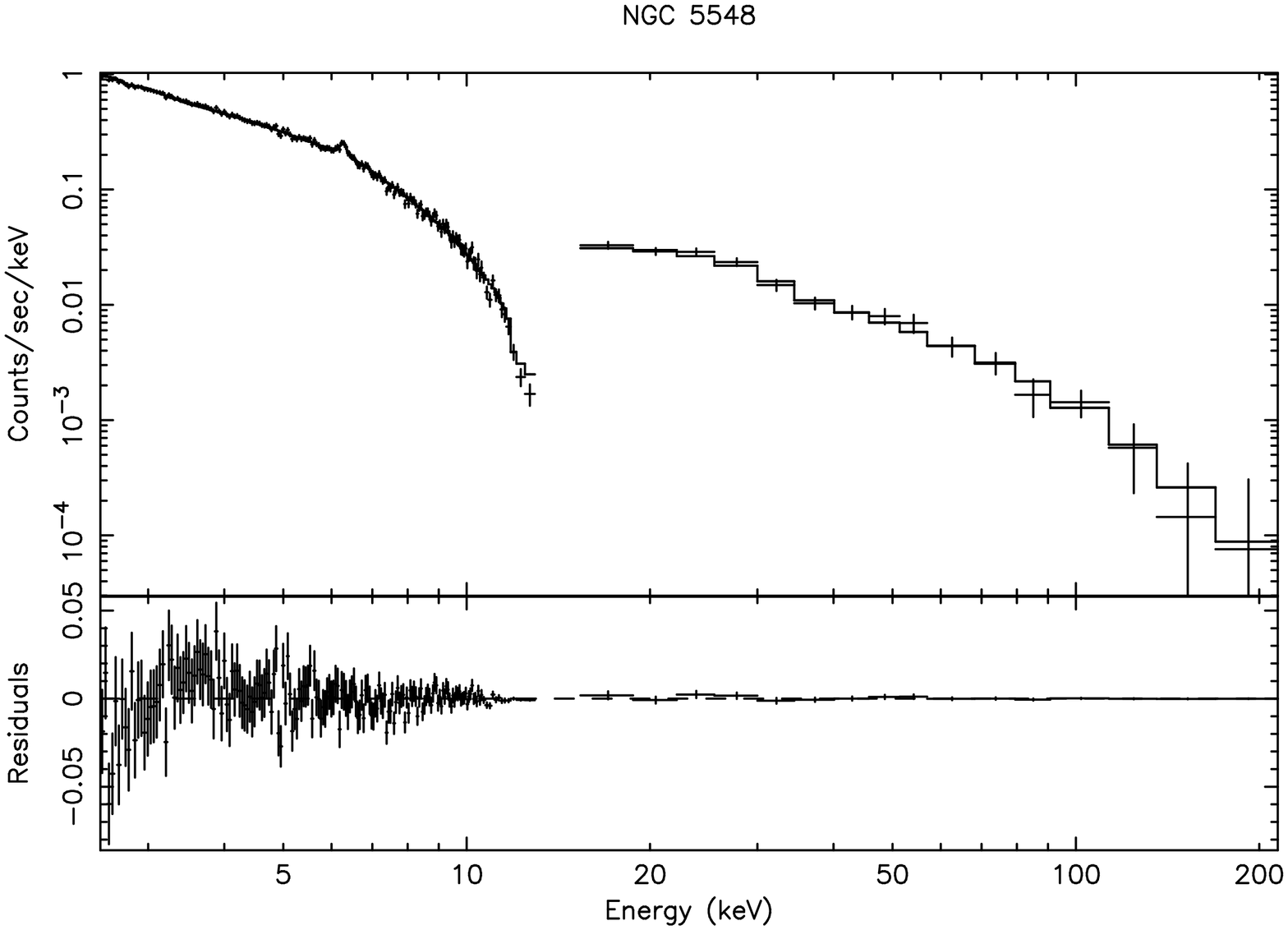,width=5.6cm,angle=-90}
\end{center}
\end{minipage}
\medskip
\begin{minipage}[b]{0.5\textwidth}
\begin{center}
\epsfig{file=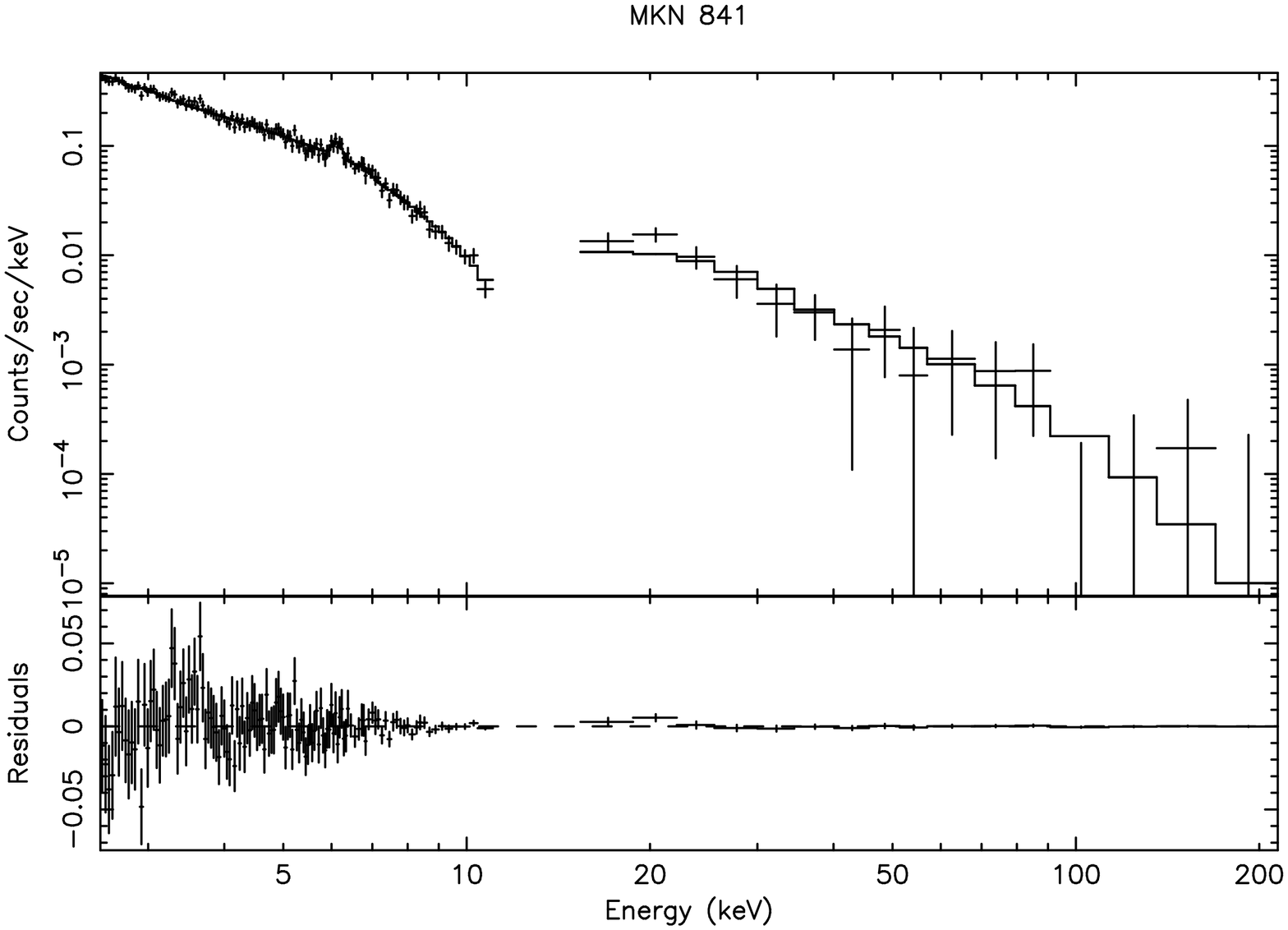,width=5.6cm,angle=-90}
\end{center}
\end{minipage}
\begin{minipage}[b]{0.5\textwidth}
\begin{center}
\epsfig{file=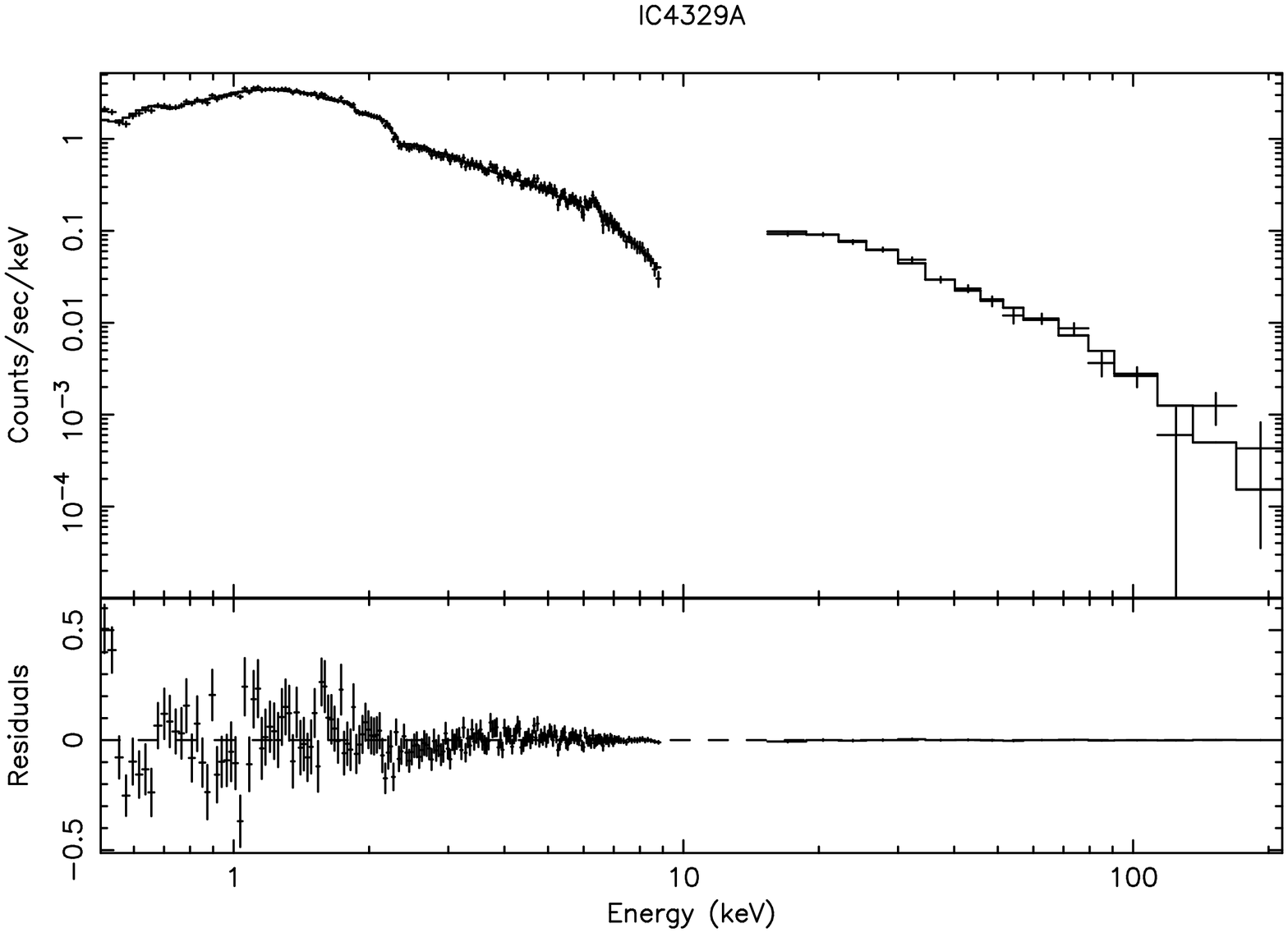,width=5.6cm,angle=-90}
\end{center}
\end{minipage}
\medskip
\begin{minipage}[b]{0.5\textwidth}
\begin{center}
\epsfig{file=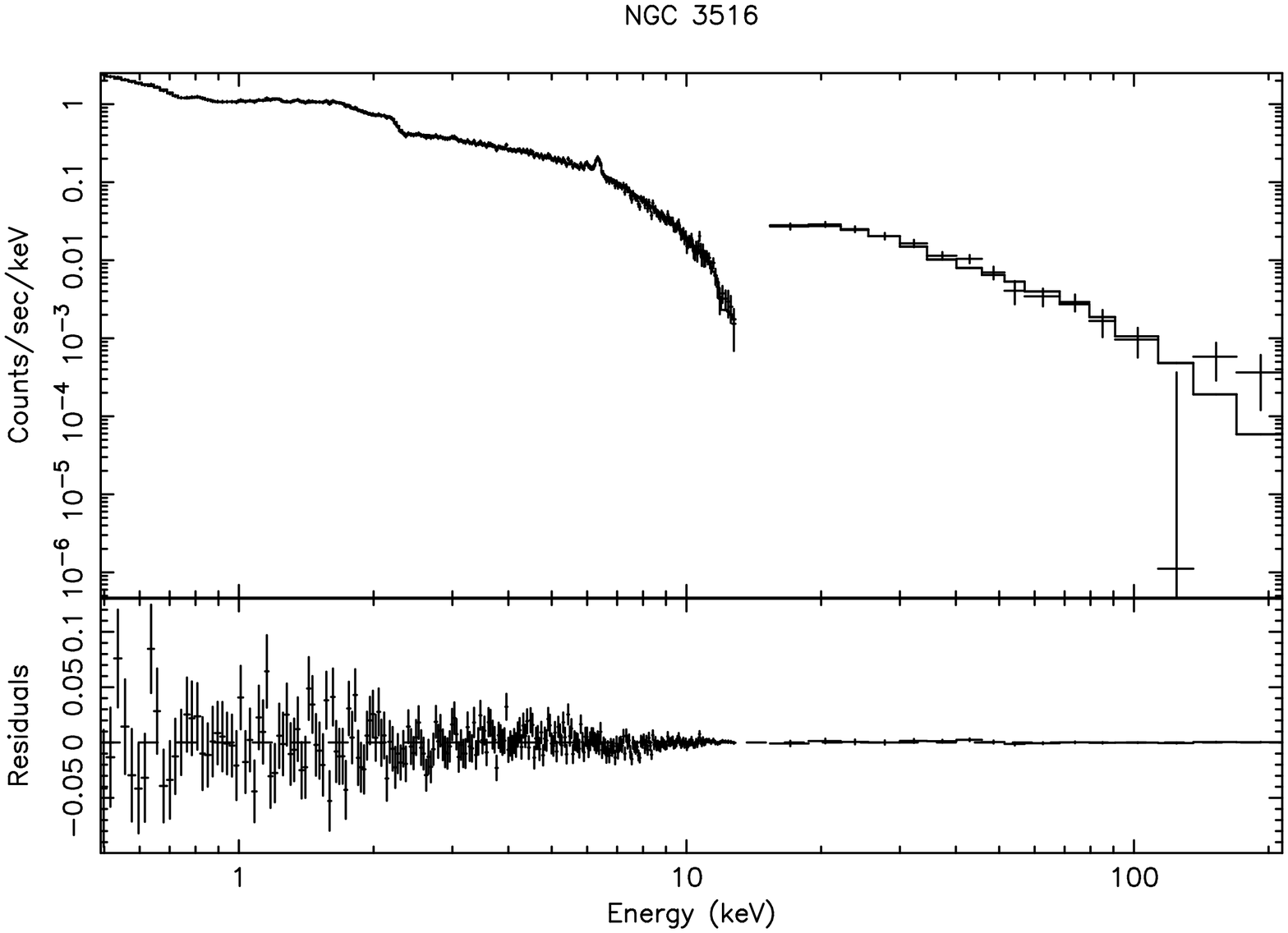,width=5.6cm,angle=-90}
\end{center}
\end{minipage}
\begin{minipage}[b]{0.5\textwidth}
\begin{center}
\epsfig{file=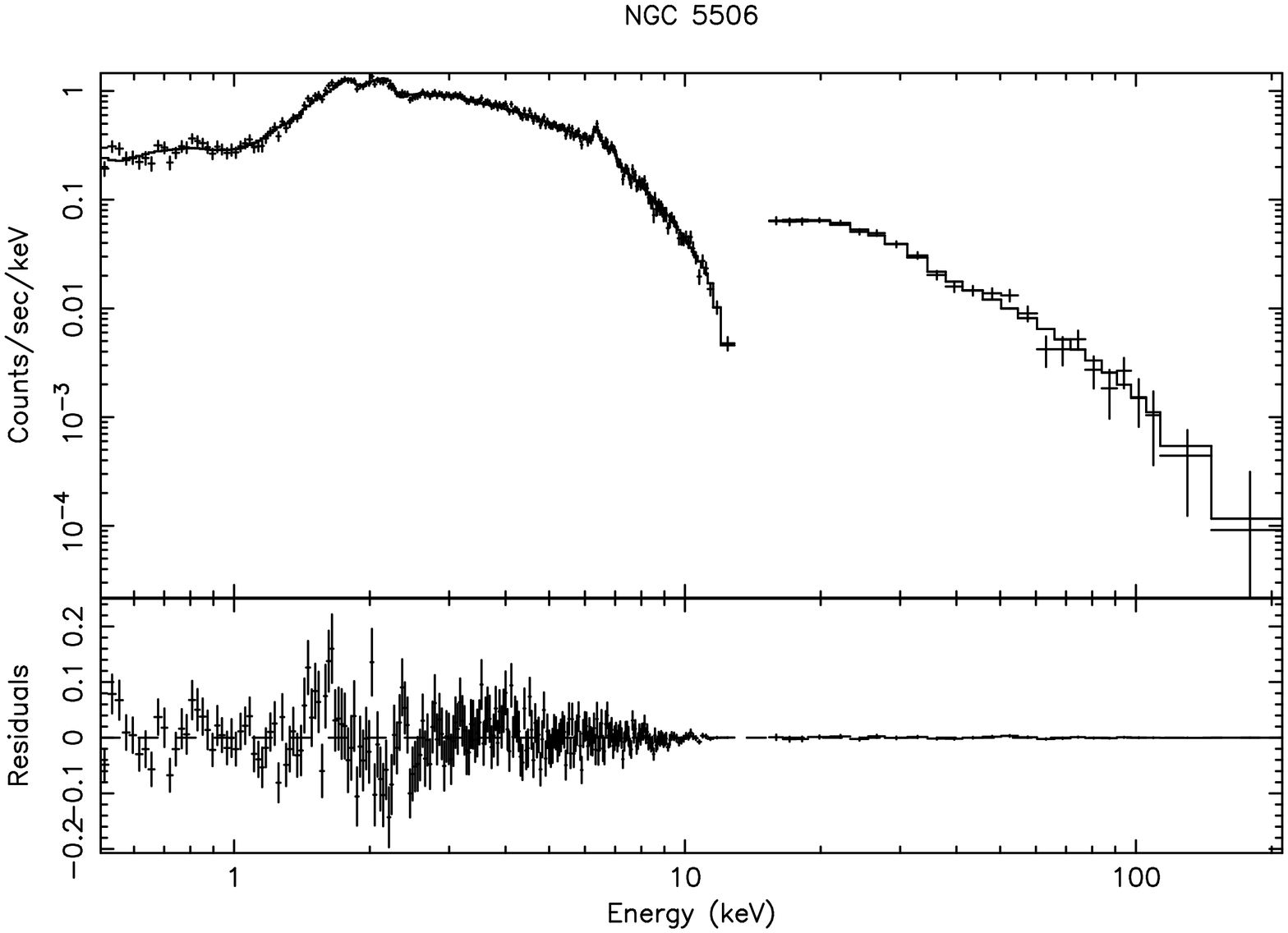,width=5.6cm,angle=-90}
\end{center}
\end{minipage}
\caption{\label{xmmsaxbms}Best fit models and residuals for all the sources of the sample. See the notes on individual objects for details.}
\end{figure*}

\subsection{\label{notes}Notes on individual sources}

\subsubsection{ESO198-G024}

The BeppoSAX and XMM-\textit{Newton} observations were analyzed separately by \citet{gua03} and \citet{por04}. The BMS gives a good $\chi^2$ (103/123 d.o.f.), but some residuals redwards the iron line are apparent in the spectrum. They can be corrected by the addition of a second narrow Gaussian line (at about 99\% confidence level), with E=$5.91^{+0.05}_{-0.06}$ keV and EW=$42^{+30}_{-22}$ eV: the parameters in Table \ref{fitnosoft} refer to this fit. The 6.4 keV line is unresolved and has a modest EW, as found by \citet{por04}. A previous XMM-\textit{Newton} observation, performed about two months before, showed a broad and more intense line \citep{gua03}, thus suggesting a variation of both profile and flux. However, it should be noted that an upper limit of 150 eV for the EW of a relativistic line is found, if added to the narrow core.

\begin{figure}[t!]
\begin{center}
\epsfig{figure=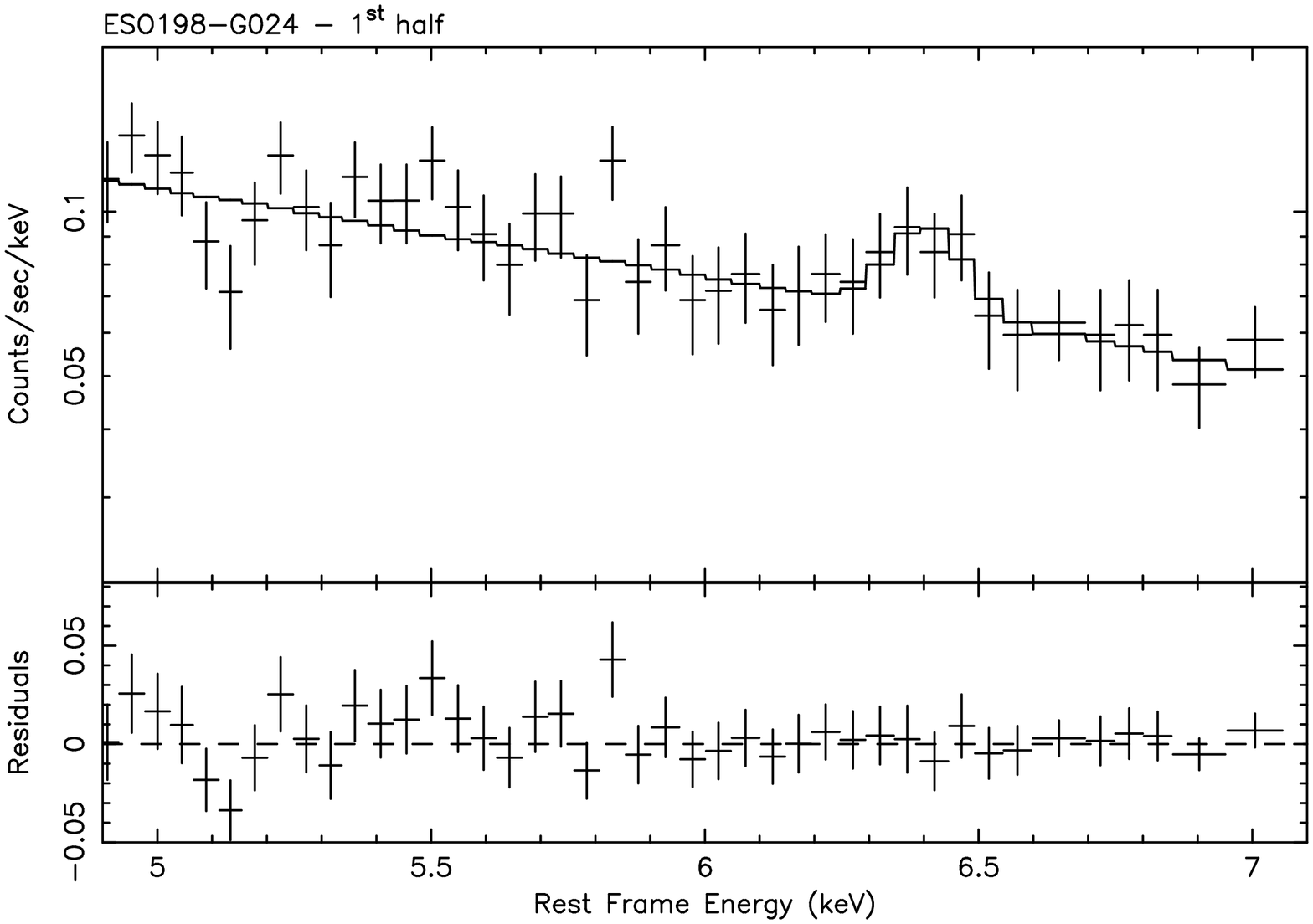, width=5.5cm, angle=-90}
\medskip
\epsfig{figure=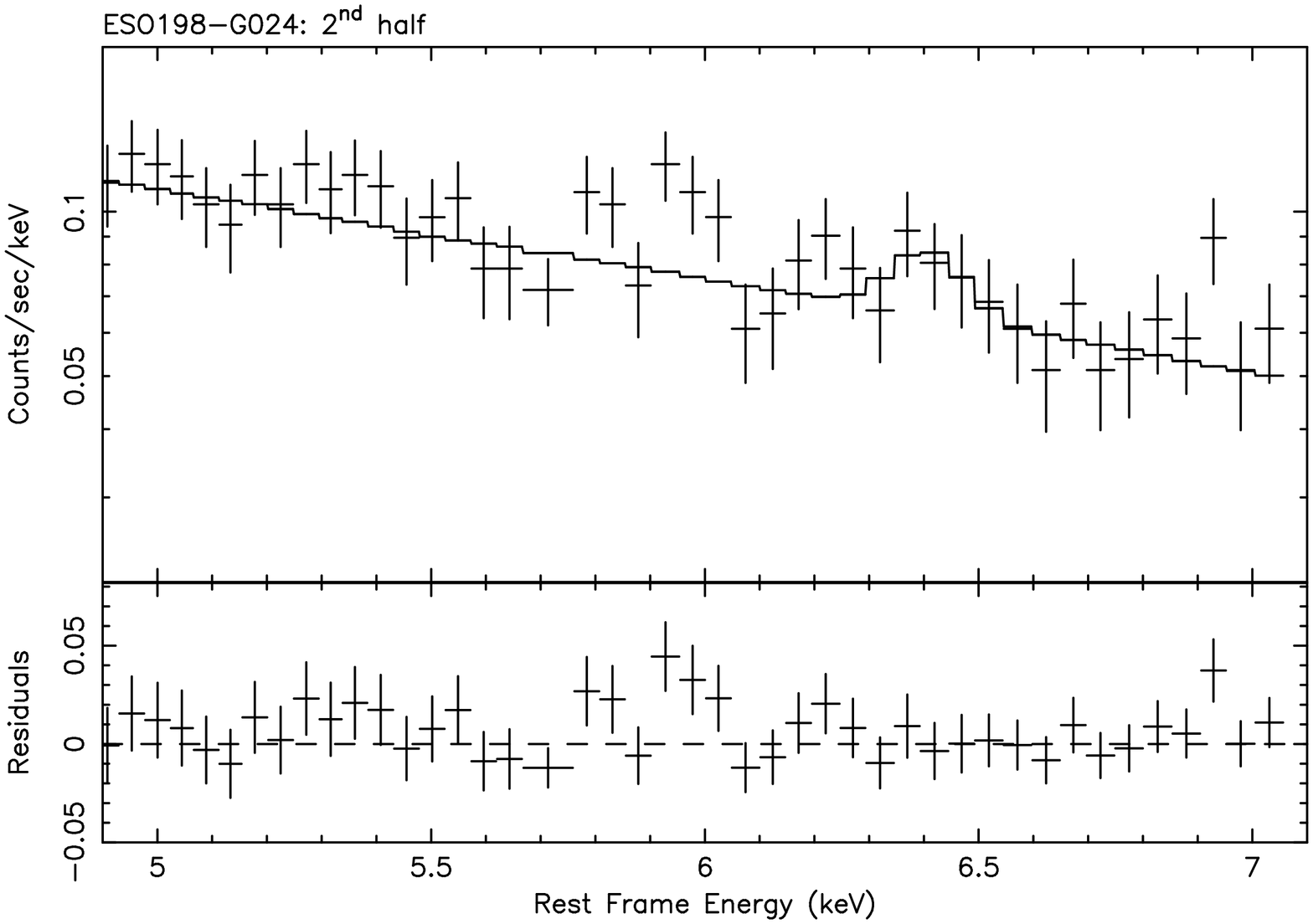,width=5.5cm,angle=-90}
\medskip
\caption{\label{esoline}ESO198-G24: EPIC pn spectra around the iron line energy for the two segments of the simultaneous observation. In addition to the 6.4 keV neutral iron line, note the presence of the feature around 6 keV only in the second half of the observation (see text for details).}
\end{center}
\end{figure}

The presence of the second feature is not reported by \citet{por04}, but \citet{gua03} found a line at $5.70^{+0.07}_{-0.12}$ keV in the other XMM-\textit{Newton} observation. We further divided the second XMM-\textit{Newton} observation in two parts of about 10 ks each. Interestingly, the feature is present (at the 98.3\% confidence level) only in the second half, at energy $5.96\pm0.05$ keV and with EW=$64^{+47}_{-36}$ eV. An upper limit of 30 eV is instead found for the first half of the observation.  Even if the statistical significance of this result is modest, it gives further support that the feature exists and is indeed varying on a short timescale. A possible interpretation for this feature consists of a 6.4 keV (rest-frame) iron line produced in a well-confined region of the accretion disk. If this is the case, the line photons suffer gravitational redshift and (relativistic) Doppler effects, as functions of the observed orbital phase, but the integrated line width would be in many cases narrow, because only the blue horn can be visible in moderate signal-to-noise ratio data. We defer the reader to \citet{dov04}, for a comprehensive discussion on this possibility and the comparison of theoretical models with data from ESO198-G024 and other sources.

\subsubsection{MCG-02-58-022}

The BMS results in a good fit ($\chi^2$=127/131). However, residuals around the iron K$\alpha$ line show that its profile is not well modelled and the line centroid is loosely constrained around 6.29 keV. Interestingly, \citet{wgy01} reported strong line variability, both in centroid and in flux, between three ASCA observations, with a value of $6.27\pm0.09$ keV in one case. Moreover, leaving the widths of the Fe K$\alpha$ and K$\beta$ lines unlinked, we get an improvement of the $\chi^2$ (at a 98\% confidence level), with two values significantly different from each other, being $\sigma=195^{+160}_{-90}$ and $<95$ eV respectively. This value for the K$\alpha$ width is consistent with the FWHM$\simeq30\,000$  km s$^{-1}$ found in the first ASCA observation \citep{weav95}.

The upper limit on the Fe K$\beta$ flux is very large, being of the order of the K$\alpha$ flux. It is then possible that there is a significant contribution from a \ion{Fe}{xxvi} line. Indeed, if the line energy is left free to vary, the above-mentioned upper limit becomes a 2 $\sigma$ detection of a $7.01^{+0.06}_{-0.05}$ keV line with an EW comparable to that of the neutral iron line.

\subsubsection{NGC 7213}

The simultaneous BeppoSAX/XMM-\textit{Newton} observation was analyzed in detail by \citet{bianchi03b}. The BMS gives an acceptable fit, but the inclusion of two narrow lines at 6.68 and 6.97 keV significantly improves it. The parameters listed in Table \ref{fitnosoft} refer to this model. Notably, the source is the only bright Seyfert 1 observed by BeppoSAX without a significant Compton reflection component \citep[see e.g.][]{per02}.

\subsubsection{NGC 5548}

This simultaneous observation was already published by \citet{pounds03}. The BMS gives an acceptable fit ($\chi^2$=206/192). The parameters, shown in Table \ref{fitnosoft}, are consistent with those found by \citet{pounds03}. In particular, the line width is marginally resolved, with a value consistent with that measured by \textit{Chandra} \citep{yaq01}. These results are in contrast with the broad line found by \textit{ASCA} \citep{nan97}. However, an analysis on eight \textit{ASCA} observations revealed a possible variability in the line flux, but no evidence for a broad line \citep{wgy01}.

\subsubsection{MKN 841}

The simultaneous observation was published by \citet{petr02}. Differently from these authors, we used in our fits a single pn spectrum, as a result of the merging of the two nearly contiguous observations (see Sect. \ref{xmmdata} for details). The BMS gives a very good fit ($\chi^2$=107/134 d.o.f.), without the inclusion of any other further component. The best fit parameters are consistent with those found by \citet{petr02}. We also note that these authors reported evidence for a flux variability on the iron line between the two contiguous observations that we merged in the present paper. We will take into account this issue in the discussion (see Sect. \ref{discussion}).

\subsubsection{IC 4329A}

The simultaneous observation was already analyzed by \citet{gond01}. However their results are affected by an incorrect value of the normalization factor between the MECS and the PDS. The BMS gives an acceptable fit ($\chi^2$=263/207 d.o.f.), with the inclusion of an edge at $0.72\pm0.02$ keV, this source being one of those whose soft X-ray spectrum has been analyzed. The line width is resolved, but its value is lower and only marginally consistent with that measured by BeppoSAX \citep{per99,per02}, while inconsistent with the one found by ASCA \citep{dmz00}.

\begin{figure}
\begin{center}
\epsfig{figure=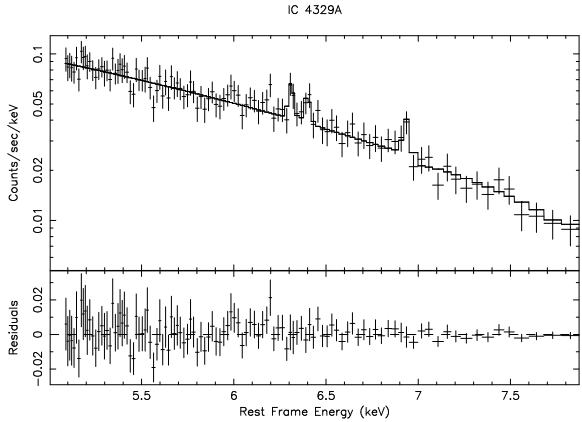,width=5.5cm,angle=-90}
\caption{\label{ic4329aheg}IC~4329A: the \textit{Chandra} HEG spectrum around the iron line.}
\end{center}
\end{figure}

A \textit{Chandra} observation, even if not simultaneous, was also analyzed to better constrain the iron line width. Indeed, the HEG spectrum reveals the presence of a double-peaked feature at $6.31^{+0.02}_{-0.01}$ and $6.40\pm0.02$ keV, with EWs of around 20 eV each  and detected at 2.5 $\sigma$ confidence level (see Fig. \ref{ic4329aheg}). The total flux of the two lines is consistent with that found by XMM-\textit{Newton} for the neutral iron line. If this is the case, the only fully resolved iron line of our sample could be the result of a blend of two lines.  The line around 6.31 keV can be one more example of relativistic features arising from an orbiting spot, as already mentioned for ESO198-G024. Deferring the reader to \citet{dov04} for details on the model, this feature could be the blue horn of a $r\simeq10$ annulus profile, provided an inclination angle slightly smaller than $30\degr$. Note, however, that this is not the only interpretation consistent with these data: \citet{my04} suggested other possibilities, such as a single broad Gaussian line or a relativistic feature arising in a nearly face-on disk. A line at $6.94^{+0.05}_{-0.06}$ keV, consistent with emission from \ion{Fe}{xxvi}, is also present in the data, with an EW of $\simeq30$ eV and at a 2 $\sigma$ confidence level. This feature was also reported by \citet{my04}.

\subsubsection{NGC 3516}

The BMS for NGC~3516 includes also the soft X-ray data, because the source is probably seen through a complex absorber \citep[see e.g.][]{cos00,gmp01} and excluding the spectrum below 2.5 keV gives erroneous results on the continuum parameters. The need to model the low energy part of the spectrum required the inclusion of a large number of components to the BMS, mainly absorption edges and emission lines. The resulting $\chi^2$ is acceptable (296/256 d.o.f.): a detailed discussion on the parameters of this fit is beyond the scopes of this work. It is worth noting, however, that the source appears absorbed by a considerable column density of ionized gas (about $1.5\times10^{22}$ cm$^{-2}$), while that from neutral gas is almost negligible (around $4\times10^{20}$ cm$^{-2}$).

In addition to a 6.4 keV narrow emission line (required at the 99.99\% confidence level according to F-test), an interesting feature in the spectrum is found at $6.08\pm0.03$ keV, whose detection is also at the 99.99\% confidence level. Similar lines were found by \citet{turner02} in another XMM-\textit{Newton} observation of NGC~3516 and a \textit{Chandra} one performed simultaneously. On the other hand, our analysis of the partly overlapping 73 ks \textit{Chandra} observation does not show any evidence for features other than the 6.4 keV line. However, the upper limit to the flux of a narrow line around 6.1 keV is consistent with the flux measured in the XMM-\textit{Newton} observation. This emission line belongs to the peculiar features possibly arising in an orbiting spot in the accretion disk, as proposed for ESO198-G024 and IC~4329A. Again, we defer the reader to \citet{dov04} for details.

\subsubsection{NGC 5506}

The simultaneous BeppoSAX/XMM-\textit{Newton} observation was analyzed in detail by \citet{matt01} and \citet{bianchi03}. We defer the reader to these papers for details on the best fit model for this source, being one of those whose spectrum below 2.5 keV was included in the analysis. We only recall, as already pointed out by the above-mentioned authors, that the inclusion of two narrow lines at 6.68 and 6.97 keV significantly improves the quality of the fit. Moreover, the neutral iron line and the Compton reflection are likely produced by a Compton-thick torus, while the source is absorbed by a Compton-thin material along the line of sight. The parameters listed in Table \ref{fitnosoft} refer to this model.

\section{\label{discussion}Discussion}

\subsection{The origin of the neutral iron line and Compton reflection}

\begin{figure*}
\begin{minipage}[b]{0.5\textwidth}
\begin{center}
\epsfig{file=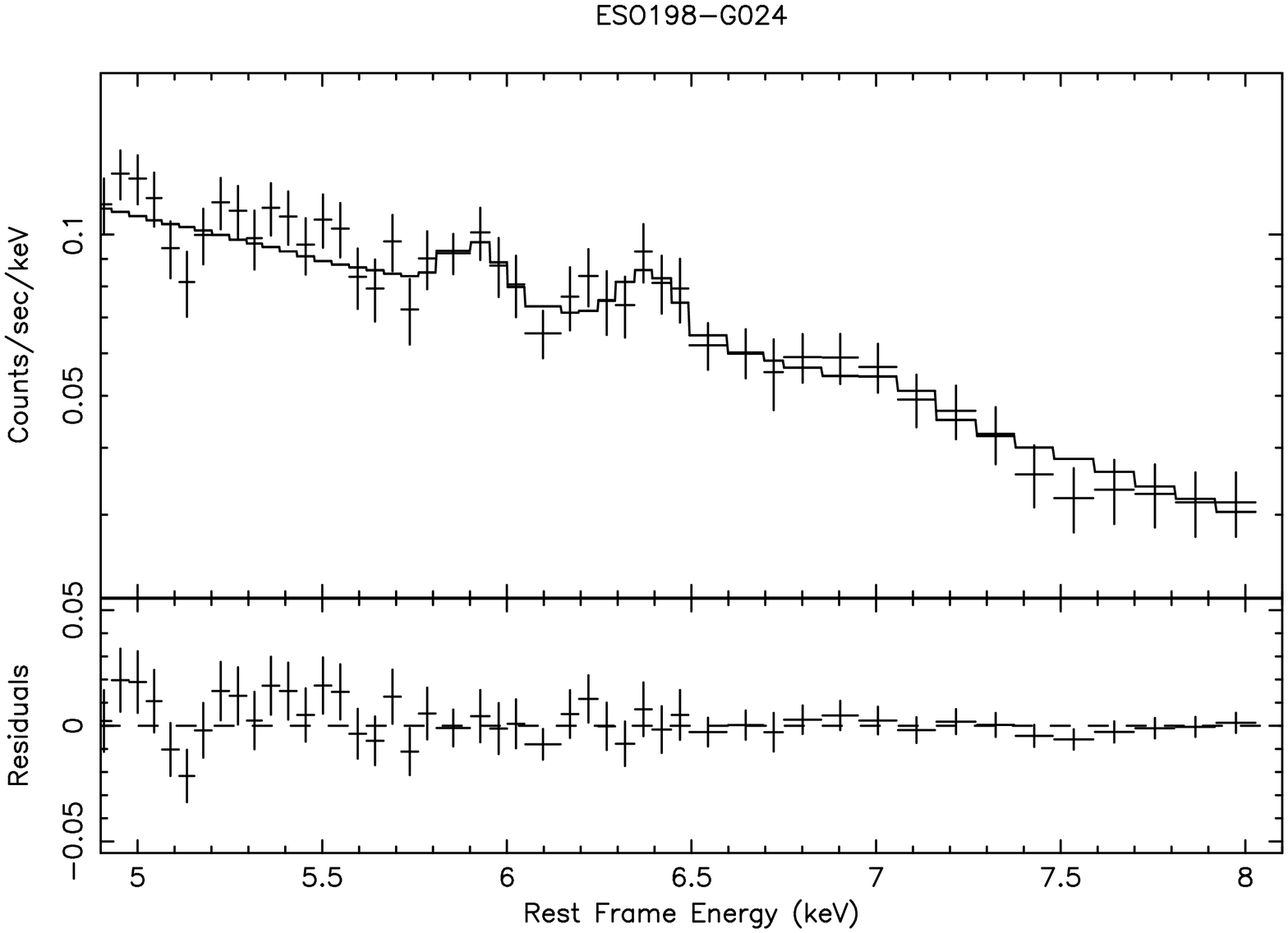,width=5.6cm,angle=-90}
\end{center}
\end{minipage}
\medskip
\begin{minipage}[b]{0.5\textwidth}
\begin{center}
\epsfig{file=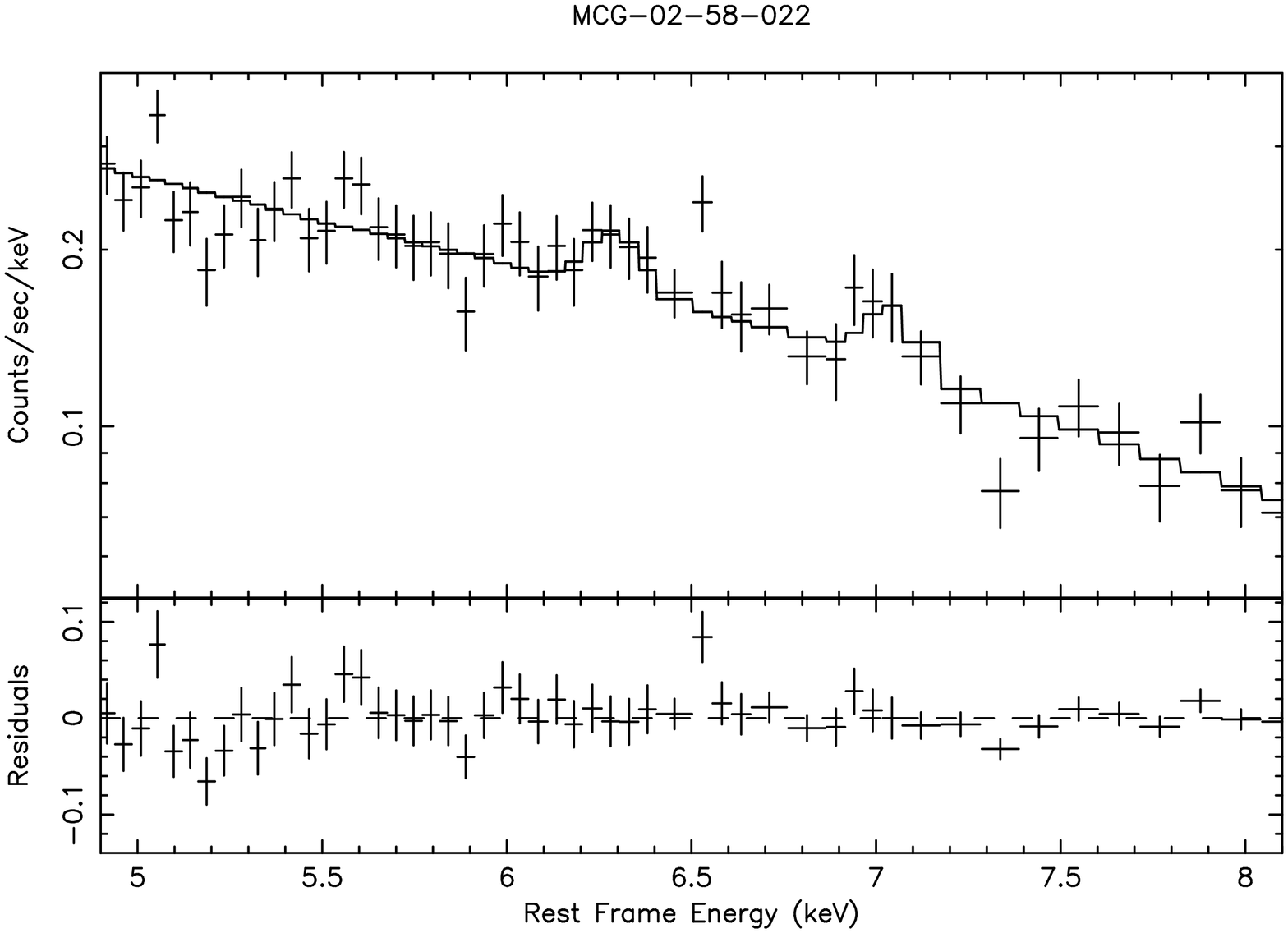,width=5.6cm,angle=-90}
\end{center}
\end{minipage}
\medskip
\begin{minipage}[b]{0.5\textwidth}
\begin{center}
\epsfig{file=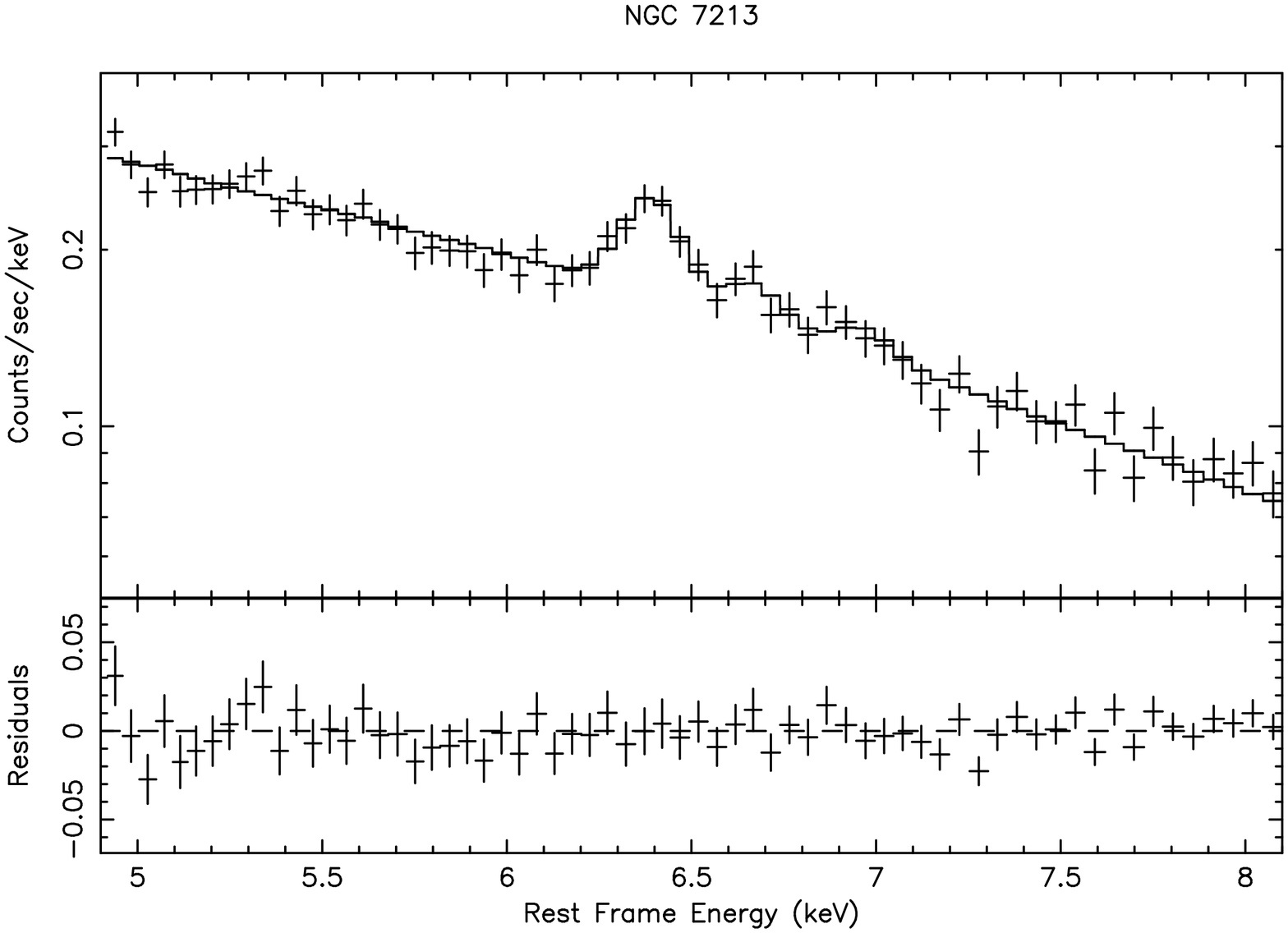,width=5.6cm,angle=-90}
\end{center}
\end{minipage}
\medskip
\begin{minipage}[b]{0.5\textwidth}
\begin{center}
\epsfig{file=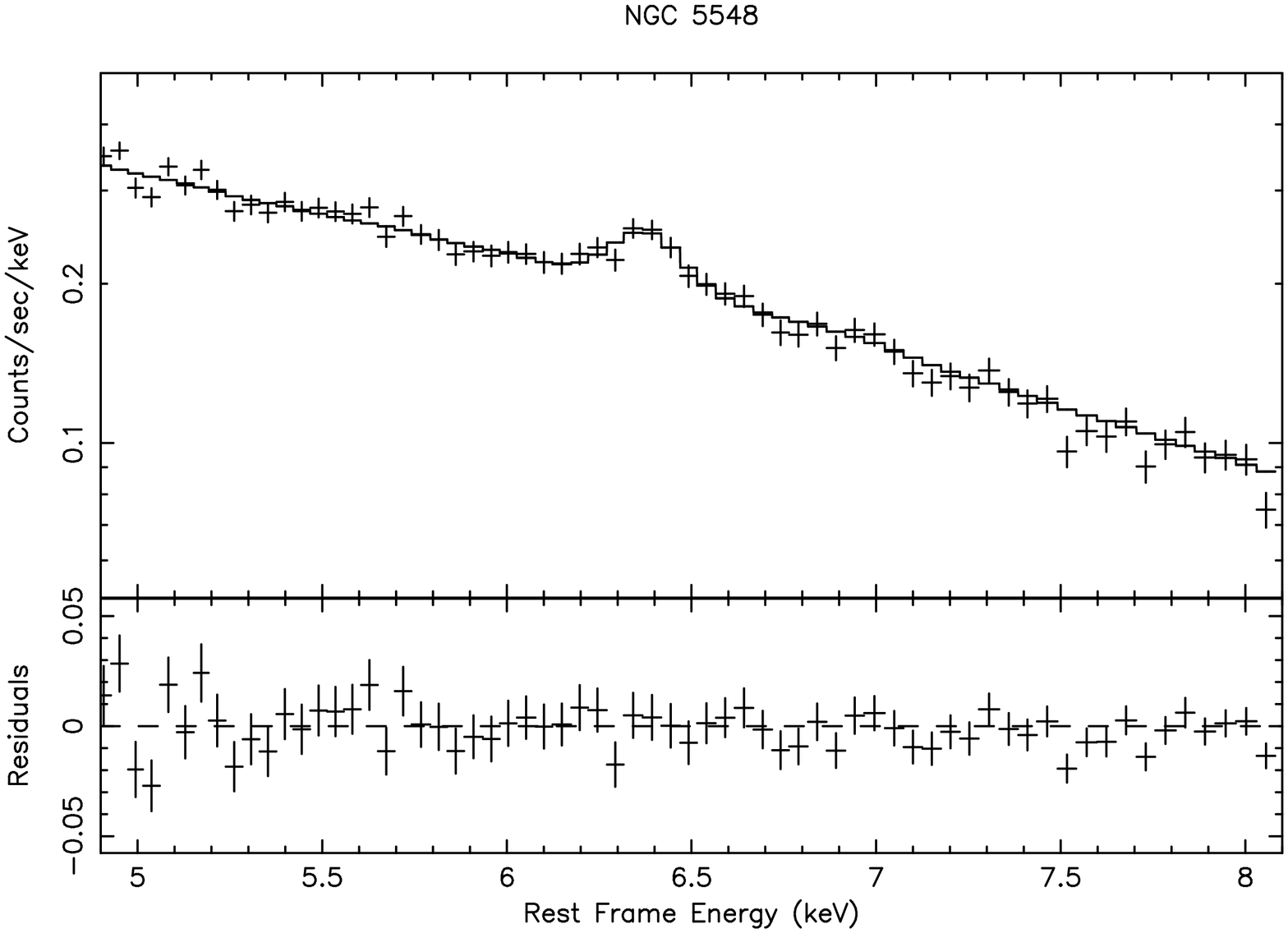,width=5.6cm,angle=-90}
\end{center}
\end{minipage}
\medskip
\begin{minipage}[b]{0.5\textwidth}
\begin{center}
\epsfig{file=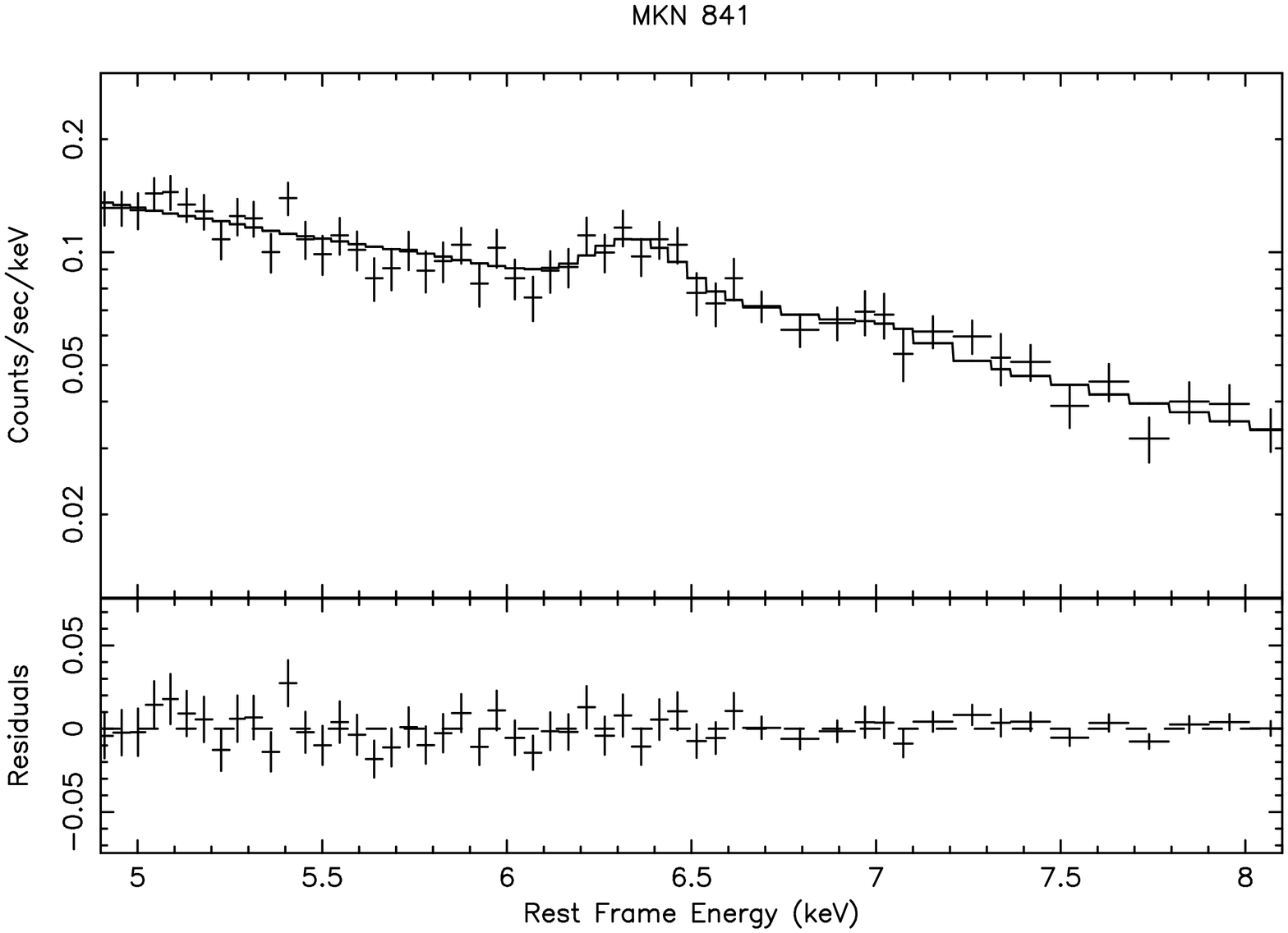,width=5.6cm,angle=-90}
\end{center}
\end{minipage}
\begin{minipage}[b]{0.5\textwidth}
\begin{center}
\epsfig{file=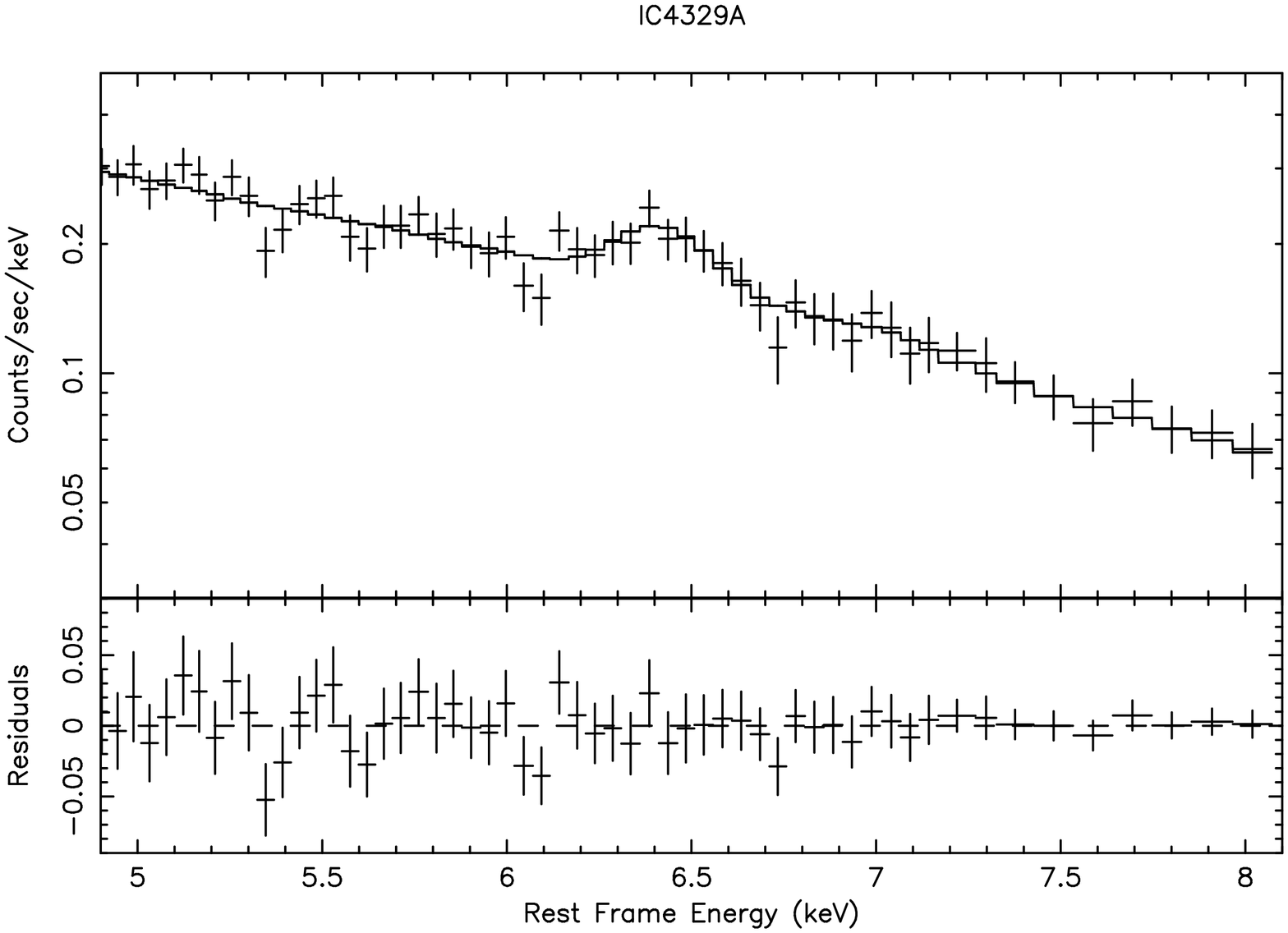,width=5.6cm,angle=-90}
\end{center}
\end{minipage}
\medskip
\begin{minipage}[b]{0.5\textwidth}
\begin{center}
\epsfig{file=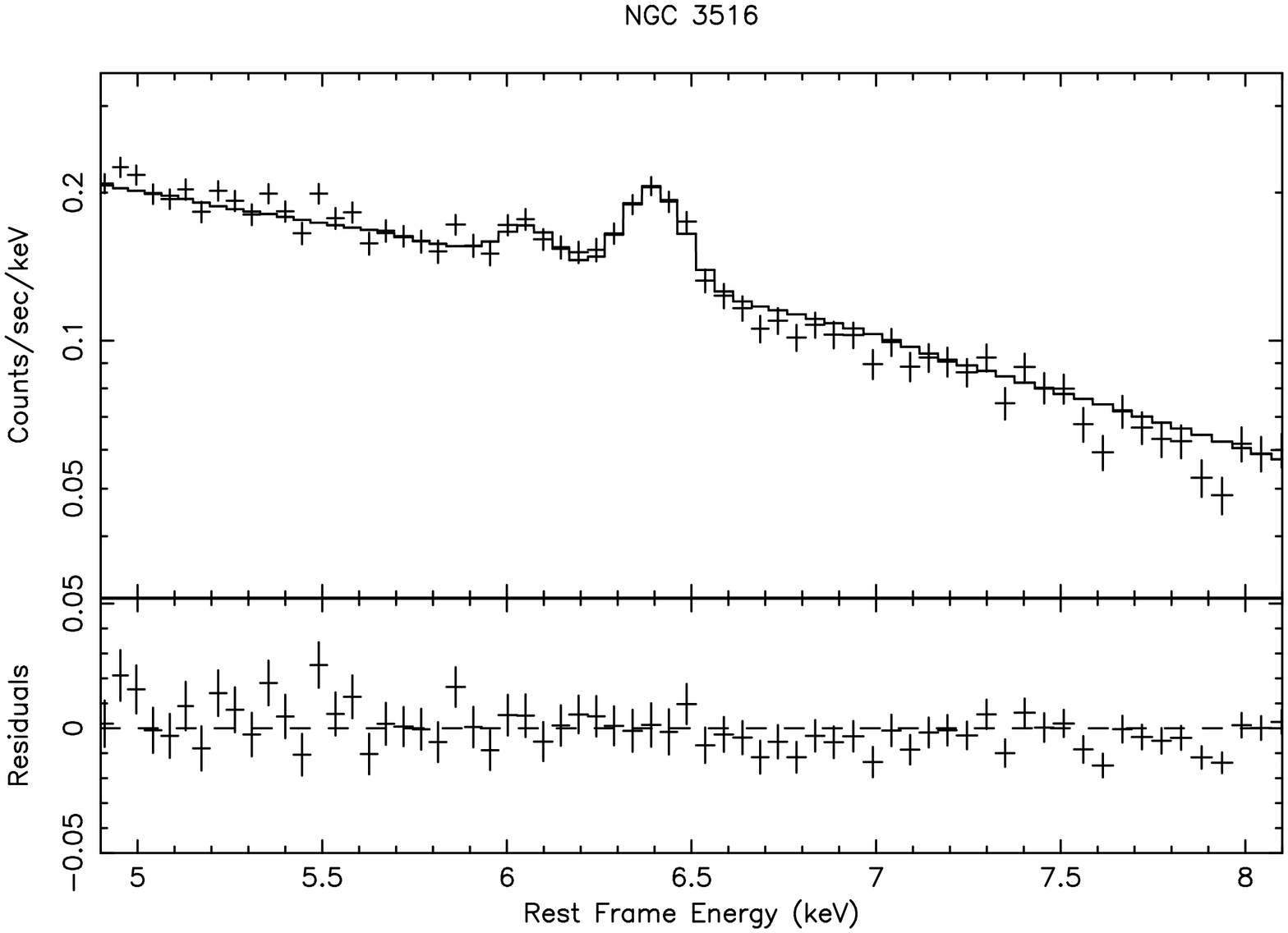,width=5.6cm,angle=-90}
\end{center}
\end{minipage}
\begin{minipage}[b]{0.5\textwidth}
\begin{center}
\epsfig{file=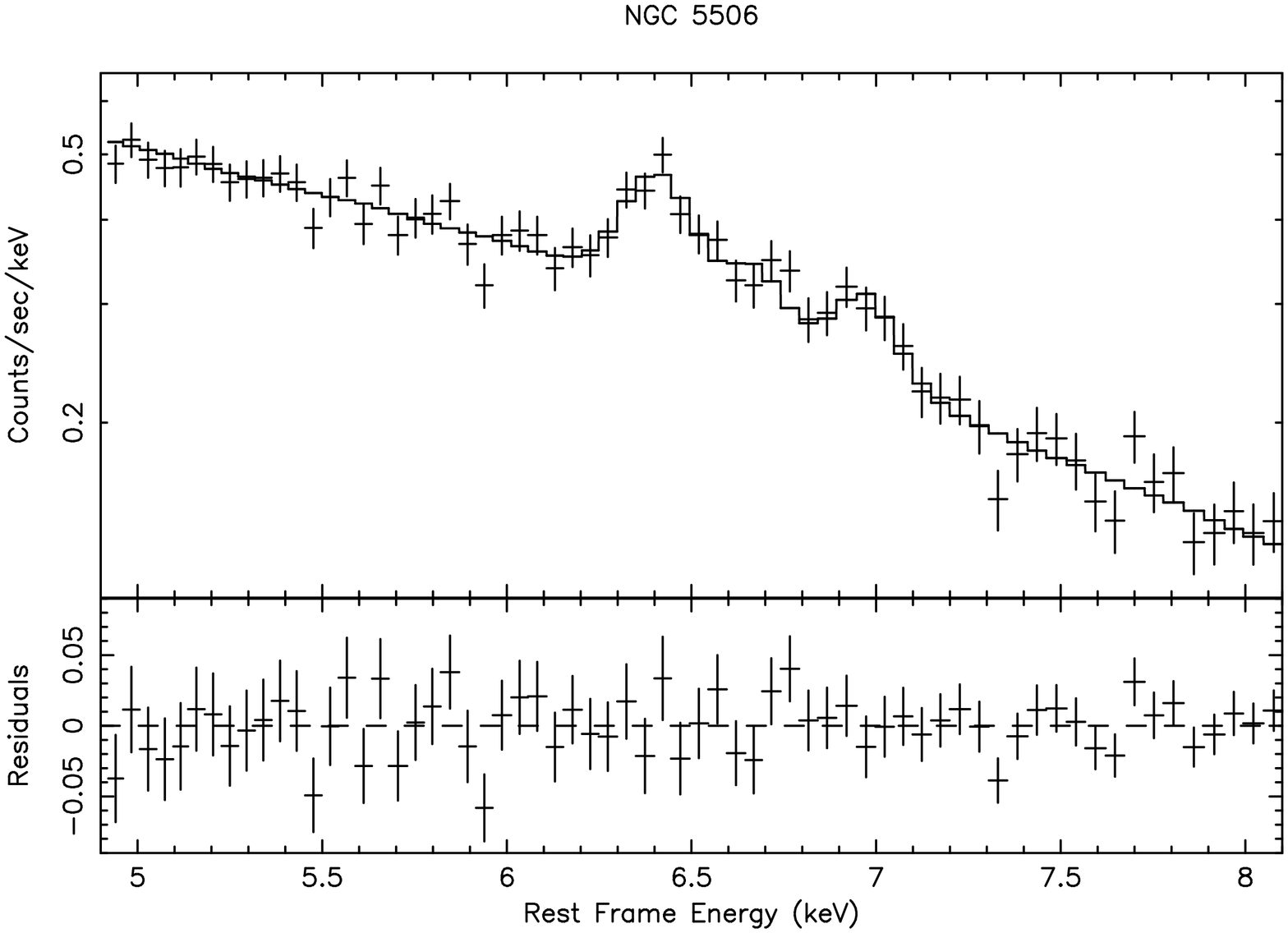,width=5.6cm,angle=-90}
\end{center}
\end{minipage}
\caption{\label{xmmsaxiron}Best fit models and residuals around the iron line energy for all the sources of the sample. Note that the spectra are corrected for the redshift of the sources. See the notes on individual objects for details.}
\end{figure*}

Inspection of Table \ref{fitnosoft} and Fig. \ref{xmmsaxiron} shows that a narrow Fe K$\alpha$ line is an ubiquitous feature in the sample. Differently from what found with ASCA \citep[see e.g.][]{nan97}, this seems now to be the rule, as already pointed out by several authors from the analysis of XMM-\textit{Newton} and \textit{Chandra} data \citep[see e.g.][and references therein]{page04,yaq04}. The most likely origin for these narrow lines is a gas located far away from the BH, such as the BLR or the torus. However, the choice between the two possibilities is not easy, since the only source with a line width resolved with high statistical significance is IC~4329A. On the other hand, even if the presence of relativistic lines is not required by the data, the statistical quality of the fits with \textsc{diskline} and \textsc{laor} models does not allow to exclude this hypothesis.

The other component present in all the sources of our sample (with the only exception of NGC~7213) is the Compton reflection produced by a fairly neutral Compton-thick material. This is independent of our choice of the inclination angle of the reflecting slab. In fact, the value of R is expected to vary roughly as the cosine of the inclination angle, resulting in variations much smaller than the errors reported in Table \ref{fitnosoft}.

These results are in principle compatible with four scenarios, which will be discussed in the following sections.

\subsubsection{The torus}

The most natural origin for both the iron line and the Compton reflection component is a Compton-thick, fairly neutral material, located far away from the BH. This description easily fits the molecular torus, whose presence in unobscured objects would confirm the predictions of Unification Models. Indeed, the values of the line EW and R are consistent with a common origin in the torus \citep[see e.g.][]{mgm03}, if we exclude the case of NGC~7213, which will be treated later. Moreover, the lack of any significant variation of the flux of the neutral iron lines during the XMM-\textit{Newton} observations (as reported in Sect. \ref{variab}) is consistent with a production in a parsec-scale material, even if the upper limits to the line variability amplitude are in general rather loose. The only exception seems to be MKN~841, since \citet{petr02} have shown that the line appears to vary on timescales as short as 10 hours. However, this result clearly needs further investigation on new data to be satisfyingly explained.

In principle this scenario could be put in question on the basis of the iron line width: a parsec-scale torus would produce a line with FWHM$\simeq760({M_8}/{r_\mathrm{pc}})^{1/2}$ km s$^{-1}$, assuming keplerian velocities \citep[see][and references therein]{yaq01}. On the other hand, the nominal best fit values for the objects where the line appears resolved in our data are around 4000-7000 km s$^{-1}$. However, the errors on these measures are so large to prevent any definitive conclusion and even \textit{Chandra} HETG observations, when available (as in the cases of NGC~3516, NGC~5506 and NGC~5548), are of little help, because those values are only marginally above the instrument resolution. The only exception is IC~4329A, which has the largest Gaussian $\sigma$ of our sample, corresponding to a FWHM= $11\,500^{+5900}_{-6500}$ km s$^{-1}$. This value is significantly larger than the one expected from the torus, while being consistent with the values of the optical/UV lines produced by the BLR observed in this source \citep{pasto79}. In this case, a significant fraction of the iron line flux must be produced in the BLR. Interestingly, IC~4329A presents the smallest value for R (if we don't consider NGC~7213) and the largest iron EW in the sample. However, all these considerations are not valid if the line width in IC~4329A is really the result of the blend of two unresolved lines, as suggested by our analysis of the \textit{Chandra} HETG data (see the individual notes on this source).

NGC~7213 is the only source which cannot easily fit this scenario. Indeed, an EW of $\simeq$ 80 eV is very difficult to reconcile with an upper limit to the Compton reflection amount as small as 0.19 \citep[see also][for the robustness of this value]{bianchi03b}, if produced by a Compton-thick torus. A possible solution is represented by a Compton-thin torus: \citet{mgm03} have shown that the shape of the continuum reflected from a Compton-thin material is rather different from the Compton-thick case, resulting in iron line EWs of several tens of eV, while R remains effectively undetectable. 

The Compton-thick torus scenario is therefore quite successful in explaining the spectra in our sample. However, it must be noted that in this picture no clear evidence for the presence of the accretion disk can be found. This is also supported by the very tight upper limits on the EWs of the relativistic iron lines when added to the BMS (Sect. \ref{models} and Table \ref{fitnosoft}). As there is no viable alternative at present to the fundamental role played by accretion disks in AGN and all the models proposed to produce the X-ray continuum necessarily imply the reprocessing of the primary emission from the disk itself, a way to reconcile data to theoretical models is likely to be found in the geometrical and physical properties of the gas that constitutes the disk. A first possibility is that the disk is highly ionized, so that most of the iron is completely stripped off, effectively suppressing the production of the line. In this case, the shape of the Compton reflection would be rather different from the one expected from neutral matter, being harder to detect and thus leaving no spectral signatures from the accretion disk. Moreover, under certain assumptions, \citet{brf01} have indeed shown that it is natural not to observe the iron line at all, in realistic models accounting for the number density of the disk atmosphere to decrease according to hydrostatic equilibrium, instead of adopting a classical constant density. A third solution would be that the accretion disk is seen almost edge-on, making the reprocessed components hardly detectable because of their low intensity. This occurrence, even if not impossible, seems implausible because it would require that most Seyfert 1s have an accretion disk with very high inclination angles, contrary to expectations from Unification Models.

\subsubsection{A `standard' accretion disk}

Even if the BMS fit is perfectly acceptable for all the sources and, in particular, the parameterization of the iron line as a Gaussian line seems correct, we decided to test an alternative model, where the narrow iron line is replaced by a relativistically broadened feature. In this scenario, no need for a Compton-thick torus is invoked, both the line and the Compton reflection component being produced by a standard accretion disk. Details on the fit parameters for this model are given in Sect. \ref{models} and results in Table \ref{fitdiskline}. The fits for this model are statistically equivalent to the BMS in all cases. However, a very large outer radius is always required by the data. This means basically (given the value of -2 adopted for the emissivity index) that most of the line is produced in the outer parts of the disk and is only marginally affected by broadening mechanisms. The resulting line profile, not surprisingly,  is not very different from a narrow Gaussian when observed at the resolution/sensitivity of present instruments. Moreover, this solution seems implausible in most sources, since the best fit outer radii are generally much larger than $1\,000$ r$_\mathrm{g}$, which is a typical value derived from theoretical models of self-gravitating disks \citep[see e.g.][and references therein]{ln89,khs04}.

A second possibility may be that the accretion disk is truncated, i.e. the inner radius is larger than the last stable orbit. This has been supported to be the case in AGN, on the base of pseudo-Newtonian hydrodynamic simulations which take fully into account radiative cooling \citep{hc00}. It was shown that, for typical accretion rates of Seyfert galaxies, truncation radii may be at 10-15 r$_\mathrm{g}$, mainly depending on the BH spin \citep{mc04}. The net effect of a truncated disk is a much less pronounced red wing for the iron line. As expected, adopting larger inner radii it is possible to have fits which are statistically equivalent to the previous ones, because the resulting line profile is again quite similar to a narrow Gaussian at the quality of our data. However, the presence, in some sources of our sample, of narrow features possibly arising from the inner radii of the accretion disk (see Sect. \ref{notes} and \ref{peculiar}) seems to exclude this hypothesis.

\subsubsection{BLR and mildly ionized disk}

In five of the sources of our sample, the iron line widths are nominally resolved, of the order of thousands of km s$^{-1}$. Even if, as already discussed, these values are rather marginal due to the large errors, it must be noted that all of them are consistent with the values of the optical/UV lines produced in the BLR, as measured in each object. Therefore, present data does not allow to exclude the possibility that all the iron lines of our sample are indeed produced in the BLR, but it would be possible in the near future, when the higher energy resolution of \textit{ASTRO}-E2 is available. Since the BLR is generally believed to be Compton-thin, a Compton-thick material is still required by the data, in order to produce the observed reflection component. However, such material should not produce a significant iron line, because, as already pointed out, no further components to the line are present in the data.

A possible solution is that the accretion disk has a degree of ionization (corresponding to \ion{Fe}{xvii}-\ion{Fe}{xxiii}) such as to prevent iron line emission, because of resonant trapping \citep[see e.g.][]{mfr93,mfr96}. At these ionization states, the Compton reflection component is still quite similar in shape and relative amount to the neutral case. In principle, this solution can be applied also to the torus, but it is difficult to conceive such an highly ionized material so far from the BH. This scenario would be a possible explanation for individual sources, but requires a fine-tuned ionization structure for the disk to be the solution for all the objects of our sample.

\subsubsection{Kerr BH and BLR}

If the iron lines are produced in the BLR, another scenario is possible, in which the Compton reflection is produced by a standard accretion disk, but the central BH has a large angular momentum. This scenario was tested adding another line component (model \textsc{laor}) to the BMS, as reported in Sect. \ref{models}. The values of the $\chi^2$ obtained with this model are statistically equivalent to that of the BMS for most sources and even better in some cases, with EWs for the relativistic line of the order of 100 eV generally not excluded by the data (see Table \ref{laor}). However, most sources require outer radii as small as 3 r$_\mathrm{g}$, thus producing a very broadened line gravitationally shifted to energies lower than 5 keV, becoming hardly distinguishable from the underlying continuum. The same considerations hold true if the outer radius is fixed to 400 r$_\mathrm{g}$, while leaving the emissivity law index free to vary. The resulting fits are as good as the previous ones, the values of $\beta$ being very large (see Table \ref{laor}): this means that only the inner radii of the disk are illuminated, producing an extreme iron line profile as before. Finally, it should be remarked that, in both cases, General relativistic effects on the Compton reflection continuum become important and the adopted \textsc{pexrav} model is no more adequate. A detailed analysis, adopting a more refined model for iron line profiles and reflection continuum arising in the inner radii of an accretion disk rotating around a Kerr BH, is deferred to a future paper.

\subsection{Other issues}

\subsubsection{Iron abundance}

In principle, it would be possible to measure the iron abundance from the depth of the iron edge and the value of the line EW with respect to the amount of reflection component. The first investigation method is supplied by the \textsc{pexrav} model, whose iron abundance parameter basically relies on the iron edge depth. While kept fixed to the solar abundance \citep{ag89} in all the fits presented so far in this paper, we therefore tried to let the iron abundance vary in the BMS. The best fit values are generally consistent with the solar abundance, with the only exception of NGC~5548, where it is marginally larger than 1. However, it is still consistent with the solar abundance at the 99\% confidence level.

On the other hand, the observed iron line EWs are close to the expected ones for the measured Compton reflection amount in the case of a solar iron abundance. Unfortunately, little can be said on the basis of this comparison, because of the errors which are generally large for both parameters. Moreover, the dependence of EW on iron abundance is not strong, being roughly logarithmic \citep{mfr97}.

\subsubsection{Fe K$\beta$ to Fe K$\alpha$ ratio}

The presence of a Fe K$\beta$ line at 7.06 keV is not statistically required in any of the spectra.  However, the upper limits to its flux are usually consistent with the expected ratio to the Fe K$\alpha$, which is about 0.16 \citep[see][]{mbm03}. In only two sources, namely NGC~7213 and NGC~5548, those upper limits are slightly lower than the theoretical value, being respectively 0.11 and 0.12.  The reason for this discrepancy could be that the iron is mildly ionized. The above-mentioned ratio of 0.16 is calculated for neutral iron and decreases with the ionization stage up to \ion{Fe}{xvii}, where the lack of M electrons prevents the emission of a K$\beta$ photon \citep{km93}. Therefore, in the case of NGC~7213 and NGC~5548 lines could be produced by iron more ionized than \ion{}{x}, following the calculations adopted by \citet{mbm03}. On the other hand, the K$\alpha$ centroid energy for these two sources are not compatible (at the 90\% confidence level) with iron more ionized than \ion{}{xii} \citep{house69}.

\subsubsection{Ionized iron lines}

The XMM-\textit{Newton} spectra of two sources, NGC~5506 and NGC~7213, require the presence of two narrow lines at 6.68 and 6.97 keV, likely produced by \ion{Fe}{xxv} and \ion{Fe}{xxvi}. An emission line at $6.94^{+0.05}_{-0.06}$ keV is also detected in the \textit{Chandra} spectrum of IC~4329A, while a similar feature at $7.01^{+0.06}_{-0.05}$ keV is present in the EPIC pn spectrum of MCG-02-58-022. The EWs of these lines are a few tens of eV, while their width is unresolved.

\citet{bm02} have shown that such lines can be produced by a Compton-thin photoionized gas illuminated by the primary continuum. The required column densities are in the range $10^{22}-10^{23}$ cm$^{-2}$, the relative strength between the \ion{Fe}{xxv} and \ion{Fe}{xxvi} lines depending on the ionization parameter of the gas. In the framework of the Unification Model, the presence of this gas in Seyfert 1s is not surprising, as  ionized iron lines have been observed in the spectra of many Seyfert 2s. In Seyfert 1s spectra, these features are clearly much diluted by the primary emission and had to await the large effective area of XMM-\textit{Newton} or, alternatively, the high resolution of \textit{Chandra} gratings to be revealed. The presence of these lines only in some sources of our sample is also not surprising, since an unlikely fine-tuning of the ionization structure of the photoionized gas would be required to produce features with an intensity large enough to be detectable above the primary continuum for any source.

Interestingly, \citet{bianchi03} have shown that in NGC~5506 the material producing the ionized iron lines is likely to be associated with a soft X-ray emitting region, which a \textit{Chandra} observation has shown to be extended on a scale of 350 pc.

\subsubsection{\label{peculiar}Peculiar narrow features}

The spectra in Fig. \ref{xmmsaxiron} clearly show narrow features at energies around 5-6 keV in NGC~3516 and ESO198-G024. Their presence was already claimed by \citet{turner02} and \citet{gua03}, respectively, and was recently followed by the observation of similar lines in NGC~7314 \citep{yaq03} and Mrk~766 \citep{tkr04}. Furthermore, our analysis of a \textit{Chandra} observation of IC~4329A suggests a possible evidence of one more feature of this kind (see Sect. \ref{notes}). We have already discussed about the possible origin of these lines in the individual notes of these sources, in terms of a 6.4 keV (rest-frame) iron line produced in a well-confined region of the accretion disk. While we defer the reader to \citet{dov04} for a very comprehensive discussion on this interpretation (and a list of possible alternatives), we would like to briefly stress some implications on our scenarios. Indeed, this identification as emission from hot orbiting spots above the disk implies that the accretion disk extends down to the last stable orbit, as required comparing the observations to the model parameters. Moreover, the ionization structure of the disk should be such as to allow iron line emission. In other words, the accretion disk may be standard with respect to the geometrical and physical properties, while the illumination could be very different from what usually assumed. In particular, as suggested by these features, it may be very anisotropic. This would be another explanation for the lack of any `classical' signature of the accretion disk in our sample.

\section{\label{conclusions}Conclusions}

We selected a sample of eight sources observed simultaneously by XMM-\textit{Newton} and BeppoSAX, taking advantage of the complementary characteristics of the two satellites, which make them one of the most effective ways to investigate the origin of iron lines. The main results of our analysis can be summarized as follows.

\begin{itemize}

\item \textbf{Narrow neutral iron lines are confirmed to be an ubiquitous component in Seyfert spectra.} Their width is generally unresolved, both at the EPIC pn and the \textit{Chandra} gratings resolutions, thus requiring a production in a material much farther from the BH than the accretion disk. Even when marginally resolved, the measured widths are typically consistent either with an origin from the torus or the BLR. This result is in agreement with the analysis of larger XMM-\textit{Newton} and \textit{Chandra} samples \citep[see e.g.][and references therein]{page04,yaq04}.

\item \textbf{None of the analyzed sources shows a broad relativistic iron line.} However, a very broad and gravitationally shifted iron line (as expected from the inner radii of a rapidly rotating BH) is not formally excluded in any of the sources.

\item \textbf{All the sources of our sample (with a single exception) show the presence of a Compton reflection component.} It is very appealing to associate the observed Compton reflection with the narrow iron line, since both features are produced by a (mostly) neutral gas and the values of R and the EW are compatible in all sources with a common origin in the same Compton-thick material. Then the most obvious identification for this matter would be  the torus, which has all the necessary properties. Moreover, its presence also in unobscured sources is in complete agreement with Unification Models.

\item \textbf{Our sample includes the only Seyfert Galaxy observed by BeppoSAX without a Compton reflection component.} In this source, namely NGC~7213, the detected iron line must originate in a Compton-thin material, such as the BLR or a Compton-thin torus.

\item \textbf{Emission lines from ionized iron are detected in three sources: NGC~5506, NGC~7213 and IC~4329A.} These features, at 6.68 and 6.97 keV (only the latter is present in the \textit{Chandra} spectrum of IC~4329A) and with EWs of a few tens eV, are readily identified with emission from \ion{Fe}{xxv} and \ion{Fe}{xxvi}. A possible evidence for a \ion{Fe}{xxvi} line was found also for MCG-02-58-022. Such lines are often observed with large EWs in Seyfert 2s, produced by an ionized, Compton-thin material. If these emitting regions are also present in the environment of Seyfert 1s, as expected in unification schemes, we should in principle be able to observe these lines as well, although with a much smaller EW due to dilution by the direct nuclear continuum. \citet{bm02} found that the resulting EWs can be as large as a few tens eV, as observed in our sources. The limited sensitivity and energy resolution of past X-ray missions prevented, until a couple of years ago, the unambiguous detection of these features in unobscured sources. The large sensitivity of XMM-\textit{Newton} allows now to detect for the first time these lines.

\item \textbf{Emission lines around 5-6 keV are detected in two sources: ESO198-G024 and NGC~3516.} These lines, even if sometimes detected with modest significance, appear to be variable on short (ESO198-G024) and long timescales (NGC~3516). A similar feature was also possibly found in the \textit{Chandra} spectrum of IC~4329A. An interpretation for these lines consists of a 6.4 keV (rest-frame) iron line produced in a well-confined region of the accretion disk. If this is the case, the line photons suffer gravitational redshift and (relativistic) Doppler effects, as for the classical relativistic profile, but the observed line width would be in many cases narrow, because only the blue horn can be visible in moderate signal-to-noise ratio data. Such a possibility would arise if the illumination of the disk is provided in some very anisotropic way, differently from the standard corona model. An example would be a very localized `hot spot' just above the disk, possibly produced by a magnetic flare. Thus, the illumination would interest just a small region of the disk, and iron emission would be produced only there. We defer the reader to \citet{dov04} for details on this model, which, if correctly explains the features, can provide a precise tool for measuring the BH mass.

\end{itemize}

The picture emerging from the results presented above strongly requires some corrections for the classical model of reprocessing from the accretion disk. On one hand, the lack of broad, relativistic lines suggests that the physical properties of the disk are likely different from those generally invoked. In particular, it is possible that the disk is truncated at radii larger than the last stable orbit, or has a degree of ionization such as to prevent iron line emission. However, it is also possible that the lines are so broad (as expected, for example, if emitted from the innermost radii of an accretion disk around a rapidly rotating BH) that their profile cannot be easily disentangled from the underlying continuum in data with present statistical quality. On the other hand, the narrow features detected around 5-6 keV, if interpreted in terms of emission from hot orbiting spots above the disk, imply that the disk indeed extends down to the last stable orbit, emitting iron lines. These contradictory results may be telling us that the illuminating properties are very complex, and still not understood.

A powerful test for all the scenarios proposed in this paper would probably be represented by high resolution time-resolved analysis, which adds another valuable piece of information to the data. However, this opportunity has to await the next generation of X-ray satellites, such as \textit{Constellation-X} and \textit{XEUS}, to be fully exploited.

\acknowledgement

We would like to thank the anonymous referee for useful suggestions. This paper is based partly on observations obtained with XMM-\textit{Newton}, an ESA science mission with instruments and contributions directly funded by ESA Member States and the USA (NASA). SB, GM, IB and GCP acknowledge ASI and MIUR (under grant \textsc{cofin-03-02-23}) for financial support.

\bibliographystyle{aa}
\bibliography{sbs}

\begin{thebibliography}{54}
\expandafter\ifx\csname natexlab\endcsname\relax\def\natexlab#1{#1}\fi

\bibitem[{{Anders} \& {Grevesse}(1989)}]{ag89}
{Anders}, E. \& {Grevesse}, N. 1989, \gca, 53, 197

\bibitem[{{Antonucci}(1993)}]{antonucci93}
{Antonucci}, R. 1993, \araa, 31, 473

\bibitem[{{Ballantyne} {et~al.}(2001){Ballantyne}, {Ross}, \& {Fabian}}]{brf01}
{Ballantyne}, D.~R., {Ross}, R.~R., \& {Fabian}, A.~C. 2001, \mnras, 327, 10

\bibitem[{{Ballantyne} {et~al.}(2003){Ballantyne}, {Vaughan}, \&
  {Fabian}}]{bvf03}
{Ballantyne}, D.~R., {Vaughan}, S., \& {Fabian}, A.~C. 2003, \mnras, 342, 239

\bibitem[{{Bianchi} {et~al.}(2003{\natexlab{a}}){Bianchi}, {Balestra}, {Matt},
  {Guainazzi}, \& {Perola}}]{bianchi03}
{Bianchi}, S., {Balestra}, I., {Matt}, G., {Guainazzi}, M., \& {Perola}, G.~C.
  2003{\natexlab{a}}, \aap, 402, 141

\bibitem[{{Bianchi} \& {Matt}(2002)}]{bm02}
{Bianchi}, S. \& {Matt}, G. 2002, \aap, 387, 76

\bibitem[{{Bianchi} {et~al.}(2003{\natexlab{b}}){Bianchi}, {Matt}, {Balestra},
  \& {Perola}}]{bianchi03b}
{Bianchi}, S., {Matt}, G., {Balestra}, I., \& {Perola}, G.~C.
  2003{\natexlab{b}}, \aap, 407, L21

\bibitem[{{Boella} {et~al.}(1997){Boella}, {Chiappetti}, {Conti}, {Cusumano},
  {del Sordo}, {La Rosa}, {Maccarone}, {Mineo}, {Molendi}, {Re}, {Sacco}, \&
  {Tripiciano}}]{boellamecs97}
{Boella}, G., {Chiappetti}, L., {Conti}, G., {et~al.} 1997, \aaps, 122, 327

\bibitem[{{Costantini} {et~al.}(2000){Costantini}, {Nicastro}, {Fruscione},
  {Mathur}, {Comastri}, {Elvis}, {Fiore}, {Salvini}, {Stirpe}, {Vignali},
  {Wilkes}, {O'Brien}, \& {Goad}}]{cos00}
{Costantini}, E., {Nicastro}, F., {Fruscione}, A., {et~al.} 2000, \apj, 544,
  283

\bibitem[{{Dickey} \& {Lockman}(1990)}]{dl90}
{Dickey}, J.~M. \& {Lockman}, F.~J. 1990, \araa, 28, 215

\bibitem[{{Done} {et~al.}(2000){Done}, {Madejski}, \& {{\. Z}ycki}}]{dmz00}
{Done}, C., {Madejski}, G.~M., \& {{\. Z}ycki}, P.~T. 2000, \apj, 536, 213

\bibitem[{{Dov\v{c}iak} {et~al.}(2004){Dov\v{c}iak}, {Bianchi}, {Guainazzi},
  {Karas}, \& {Matt}}]{dov04}
{Dov\v{c}iak}, M., {Bianchi}, S., {Guainazzi}, M., {Karas}, V., \& {Matt}, G.
  2004, MNRAS, in press

\bibitem[{{Fabian} {et~al.}(2000){Fabian}, {Iwasawa}, {Reynolds}, \&
  {Young}}]{fab00}
{Fabian}, A.~C., {Iwasawa}, K., {Reynolds}, C.~S., \& {Young}, A.~J. 2000,
  \pasp, 112, 1145

\bibitem[{{Fabian} {et~al.}(1989){Fabian}, {Rees}, {Stella}, \&
  {White}}]{fab89}
{Fabian}, A.~C., {Rees}, M.~J., {Stella}, L., \& {White}, N.~E. 1989, \mnras,
  238, 729

\bibitem[{{Fabian} {et~al.}(2002){Fabian}, {Vaughan}, {Nandra}, {Iwasawa},
  {Ballantyne}, {Lee}, {De Rosa}, {Turner}, \& {Young}}]{fab02}
{Fabian}, A.~C., {Vaughan}, S., {Nandra}, K., {et~al.} 2002, \mnras, 335, L1

\bibitem[{{Fiore} {et~al.}(1999){Fiore}, {Guainazzi}, \& {Grandi}}]{fiore99}
{Fiore}, F., {Guainazzi}, M., \& {Grandi}, P. 1999, SDC report
  (http://www.asdc.asi.it/bepposax/software/index.html)

\bibitem[{{Frontera} {et~al.}(1997){Frontera}, {Costa}, {dal Fiume}, {Feroci},
  {Nicastro}, {Orlandini}, {Palazzi}, \& {Zavattini}}]{fronterapds97}
{Frontera}, F., {Costa}, E., {dal Fiume}, D., {et~al.} 1997, \aaps, 122, 357

\bibitem[{{Gondoin} {et~al.}(2001){Gondoin}, {Barr}, {Lumb}, {Oosterbroek},
  {Orr}, \& {Parmar}}]{gond01}
{Gondoin}, P., {Barr}, P., {Lumb}, D., {et~al.} 2001, \aap, 378, 806

\bibitem[{{Guainazzi}(2003)}]{gua03}
{Guainazzi}, M. 2003, \aap, 401, 903

\bibitem[{{Guainazzi} {et~al.}(2001){Guainazzi}, {Marshall}, \&
  {Parmar}}]{gmp01}
{Guainazzi}, M., {Marshall}, W., \& {Parmar}, A.~N. 2001, \mnras, 323, 75

\bibitem[{{House}(1969)}]{house69}
{House}, L.~L. 1969, \apjs, 18, 21

\bibitem[{{Hujeirat} \& {Camenzind}(2000)}]{hc00}
{Hujeirat}, A. \& {Camenzind}, M. 2000, \aap, 361, L53

\bibitem[{{Kaastra} \& {Mewe}(1993)}]{km93}
{Kaastra}, J.~S. \& {Mewe}, R. 1993, \aaps, 97, 443

\bibitem[{{Karas} {et~al.}(2004){Karas}, {Hur\'e}, \& {Semer\'ak}}]{khs04}
{Karas}, V., {Hur\'e}, J.~M., \& {Semer\'ak}, O. 2004, accepted for
  pubblication in Classical and Quantum Gravity (astro-ph/0401345)

\bibitem[{{Laor}(1991)}]{laor91}
{Laor}, A. 1991, \apj, 376, 90

\bibitem[{{Laor} \& {Netzer}(1989)}]{ln89}
{Laor}, A. \& {Netzer}, H. 1989, \mnras, 238, 897

\bibitem[{{Lumb} {et~al.}(2002){Lumb}, {Warwick}, {Page}, \& {De
  Luca}}]{lumb02}
{Lumb}, D.~H., {Warwick}, R.~S., {Page}, M., \& {De Luca}, A. 2002, \aap, 389,
  93

\bibitem[{{Magdziarz} \& {Zdziarski}(1995)}]{mz95}
{Magdziarz}, P. \& {Zdziarski}, A.~A. 1995, \mnras, 273, 837

\bibitem[{{Matt} {et~al.}(1997){Matt}, {Fabian}, \& {Reynolds}}]{mfr97}
{Matt}, G., {Fabian}, A.~C., \& {Reynolds}, C.~S. 1997, \mnras, 289, 175

\bibitem[{{Matt} {et~al.}(1993){Matt}, {Fabian}, \& {Ross}}]{mfr93}
{Matt}, G., {Fabian}, A.~C., \& {Ross}, R.~R. 1993, \mnras, 262, 179

\bibitem[{{Matt} {et~al.}(1996){Matt}, {Fabian}, \& {Ross}}]{mfr96}
{Matt}, G., {Fabian}, A.~C., \& {Ross}, R.~R. 1996, \mnras, 278, 1111

\bibitem[{{Matt} {et~al.}(2003){Matt}, {Guainazzi}, \& {Maiolino}}]{mgm03}
{Matt}, G., {Guainazzi}, M., \& {Maiolino}, R. 2003, \mnras, 342, 422

\bibitem[{{Matt} {et~al.}(2001){Matt}, {Guainazzi}, {Perola}, {Fiore},
  {Nicastro}, {Cappi}, \& {Piro}}]{matt01}
{Matt}, G., {Guainazzi}, M., {Perola}, G.~C., {et~al.} 2001, \aap, 377, L31

\bibitem[{{Matt} {et~al.}(2000){Matt}, {Perola}, {Fiore}, {Guainazzi},
  {Nicastro}, \& {Piro}}]{matt00}
{Matt}, G., {Perola}, G.~C., {Fiore}, F., {et~al.} 2000, \aap, 363, 863

\bibitem[{{McKernan} \& {Yaqoob}(2004)}]{my04}
{McKernan}, B. \& {Yaqoob}, T. 2004, accepted for publication in ApJ
  (astro-ph/0311551)

\bibitem[{{Molendi} {et~al.}(2003){Molendi}, {Bianchi}, \& {Matt}}]{mbm03}
{Molendi}, S., {Bianchi}, S., \& {Matt}, G. 2003, \mnras, 343, L1

\bibitem[{{M{\"u}ller} \& {Camenzind}(2004)}]{mc04}
{M{\"u}ller}, A. \& {Camenzind}, M. 2004, \aap, 413, 861

\bibitem[{{Nandra} {et~al.}(1997){Nandra}, {George}, {Mushotzky}, {Turner}, \&
  {Yaqoob}}]{nan97}
{Nandra}, K., {George}, I.~M., {Mushotzky}, R.~F., {Turner}, T.~J., \&
  {Yaqoob}, T. 1997, \apj, 477, 602

\bibitem[{{Page} {et~al.}(2004){Page}, {O'Brien}, {Reeves}, \&
  {Turner}}]{page04}
{Page}, K.~L., {O'Brien}, P.~T., {Reeves}, J.~N., \& {Turner}, M.~J.~L. 2004,
  \mnras, 347, 316

\bibitem[{{Pastoriza}(1979)}]{pasto79}
{Pastoriza}, M.~G. 1979, \apj, 234, 837

\bibitem[{{Perola} {et~al.}(1999){Perola}, {Matt}, {Cappi}, {Dal Fiume},
  {Fiore}, {Guainazzi}, {Mineo}, {Molendi}, {Nicastro}, {Piro}, \&
  {Stirpe}}]{per99}
{Perola}, G.~C., {Matt}, G., {Cappi}, M., {et~al.} 1999, \aap, 351, 937

\bibitem[{{Perola} {et~al.}(2002){Perola}, {Matt}, {Cappi}, {Fiore},
  {Guainazzi}, {Maraschi}, {Petrucci}, \& {Piro}}]{per02}
{Perola}, G.~C., {Matt}, G., {Cappi}, M., {et~al.} 2002, \aap, 389, 802

\bibitem[{{Petrucci} {et~al.}(2002){Petrucci}, {Henri}, {Maraschi}, {Ferrando},
  {Matt}, {Mouchet}, {Perola}, {Collin}, {Dumont}, {Haardt}, \&
  {Koch-Miramond}}]{petr02}
{Petrucci}, P.~O., {Henri}, G., {Maraschi}, L., {et~al.} 2002, \aap, 388, L5

\bibitem[{{Porquet} {et~al.}(2004){Porquet}, {Kaastra}, {Page}, {O'Brien},
  {Ward}, \& {Dubau}}]{por04}
{Porquet}, D., {Kaastra}, J.~S., {Page}, K.~L., {et~al.} 2004, \aap, 413, 913

\bibitem[{{Pounds} {et~al.}(2003){Pounds}, {Reeves}, {Page}, {Edelson}, {Matt},
  \& {Perola}}]{pounds03}
{Pounds}, K.~A., {Reeves}, J.~N., {Page}, K.~L., {et~al.} 2003, \mnras, 341,
  953

\bibitem[{{Schurch} \& {Warwick}(2002)}]{sw02}
{Schurch}, N.~J. \& {Warwick}, R.~S. 2002, \mnras, 334, 811

\bibitem[{{Str{\"u}der} {et~al.}(2001){Str{\"u}der}, {Briel}, {Dennerl},
  {Hartmann}, {Kendziorra}, {Meidinger}, {Pfeffermann}, {Reppin}, {Aschenbach},
  {Bornemann}, {Br{\" a}uninger}, {Burkert}, {Elender}, {Freyberg}, {Haberl},
  {Hartner}, {Heuschmann}, {Hippmann}, {Kastelic}, {Kemmer}, {Kettenring},
  {Kink}, {Krause}, {M{\" u}ller}, {Oppitz}, {Pietsch}, {Popp}, {Predehl},
  {Read}, {Stephan}, {St{\" o}tter}, {Tr{\" u}mper}, {Holl}, {Kemmer},
  {Soltau}, {St{\" o}tter}, {Weber}, {Weichert}, {von Zanthier},
  {Carathanassis}, {Lutz}, {Richter}, {Solc}, {B{\" o}ttcher}, {Kuster},
  {Staubert}, {Abbey}, {Holland}, {Turner}, {Balasini}, {Bignami}, {La
  Palombara}, {Villa}, {Buttler}, {Gianini}, {Lain{\' e}}, {Lumb}, \&
  {Dhez}}]{struder01}
{Str{\"u}der}, L., {Briel}, U., {Dennerl}, K., {et~al.} 2001, \aap, 365, L18

\bibitem[{{Turner} {et~al.}(2004){Turner}, {Kraemer}, \& {Reeves}}]{tkr04}
{Turner}, T.~J., {Kraemer}, S.~B., \& {Reeves}, J.~N. 2004, accepted for
  publication in ApJ (astro-ph/0310885)

\bibitem[{{Turner} {et~al.}(2002){Turner}, {Mushotzky}, {Yaqoob}, {George},
  {Snowden}, {Netzer}, {Kraemer}, {Nandra}, \& {Chelouche}}]{turner02}
{Turner}, T.~J., {Mushotzky}, R.~F., {Yaqoob}, T., {et~al.} 2002, \apjl, 574,
  L123

\bibitem[{{Weaver} {et~al.}(2001){Weaver}, {Gelbord}, \& {Yaqoob}}]{wgy01}
{Weaver}, K.~A., {Gelbord}, J., \& {Yaqoob}, T. 2001, \apj, 550, 261

\bibitem[{{Weaver} {et~al.}(1995){Weaver}, {Nousek}, {Yaqoob}, {Hayashida}, \&
  {Murakami}}]{weav95}
{Weaver}, K.~A., {Nousek}, J., {Yaqoob}, T., {Hayashida}, K., \& {Murakami}, S.
  1995, \apj, 451, 147

\bibitem[{{Yaqoob} {et~al.}(2003){Yaqoob}, {George}, {Kallman}, {Padmanabhan},
  {Weaver}, \& {Turner}}]{yaq03}
{Yaqoob}, T., {George}, I.~M., {Kallman}, T.~R., {et~al.} 2003, \apj, 596, 85

\bibitem[{{Yaqoob} {et~al.}(2001){Yaqoob}, {George}, {Nandra}, {Turner},
  {Serlemitsos}, \& {Mushotzky}}]{yaq01}
{Yaqoob}, T., {George}, I.~M., {Nandra}, K., {et~al.} 2001, \apj, 546, 759

\bibitem[{{Yaqoob} \& {Padmanabhan}(2004)}]{yaq04}
{Yaqoob}, T. \& {Padmanabhan}, U. 2004, accepted for publication in ApJ
  (astro-ph/0311551)

\end{thebibliography}

\end{document}